\documentclass[sigconf, nonacm]{acmart}
\def\mode{0} % Comment this line out if you want the TR-mode

\newcommand\vldbdoi{10.14778/3476249.3476274}
\newcommand\vldbpages{2216 - 2229}
\newcommand\vldbvolume{14}
\newcommand\vldbissue{11}
\newcommand\vldbyear{2021}
\newcommand\vldbauthors{\authors}
\newcommand\vldbtitle{\shorttitle} 
\newcommand\vldbavailabilityurl{https://disc.bu.edu/lsm-compaction}
\newcommand\vldbpagestyle{empty}

\usepackage{booktabs} % For formal tables
\usepackage[ruled,vlined]{algorithm2e}
\usepackage{amsmath}
\usepackage{wrapfig} 
\usepackage{balance}
\usepackage{enumitem}
\usepackage{moresize}
\usepackage{subcaption}
\usepackage{epstopdf}
\usepackage{graphicx} 
\usepackage{marginnote}
\usepackage{color}

\usepackage{multirow} 
\usepackage{makecell}

\usepackage{enumitem}
\usepackage{bm}
\usepackage{balance}
\usepackage[normalem]{ulem}
\usepackage{float}
\usepackage[printwatermark]{xwatermark}

\usepackage{pifont}% http://ctan.org/pkg/pifont
\newcommand{\cmark}{\ding{51}}%

\definecolor{myRed}{rgb}{0.73, 0.31, 0.28}
\definecolor{myBlue}{rgb}{0, 0.44, 1}
\definecolor{myGreen}{rgb}{0.31, 0.78, 0.47}
\definecolor{myGrey}{rgb}{0.57, 0.64, 0.69}

\usepackage{array}
\newcommand{\PreserveBackslash}[1]{\let\temp=\\#1\let\\=\temp}
\newcolumntype{C}[1]{>{\PreserveBackslash\centering}p{#1}}
\newcolumntype{R}[1]{>{\PreserveBackslash\raggedleft}p{#1}}
\newcolumntype{L}[1]{>{\PreserveBackslash\raggedright}p{#1}}

\newcommand\Paragraph[1]{\vspace{0.02in}  \noindent \textbf{#1.}}

\newcommand\Paragraphit[1]{\vspace{0.02in}  \noindent \textit{#1.}}

\newcounter{observation}
\newcommand\Ob{O\addtocounter{observation}{1}\theobservation: }
\newcounter{minorobs}
\newcommand\mob{o\addtocounter{minorobs}{1}\theminorobs: }
\newcounter{takeaway}
\newcommand\TA{TA \addtocounter{takeaway}{1}\thetakeaway: }
\renewcommand\thetakeaway{\Roman{takeaway}}

\begin{document}

\ifx\mode\undefined
\title{Technical Report: Constructing and Constructing the LSM Compaction Design Space}
\else
\title{Constructing and Analyzing the LSM Compaction Design Space (Updated Version)}
\fi

%\author{Subhadeep Sarkar}
% \email{ssarkar1@bu.edu}
%\affiliation{%
%  % \institution{Boston University}
%  % \streetaddress{111 Cummington Mall}
%  % \city{Boston}
%  % \state{MA}
%  % \postcode{02215}
%}
%
%\author{Dimitris Staratzis}
% \email{dstara@bu.edu}
%\affiliation{%
%%   \institution{Boston University}
%%   \streetaddress{111 Cummington Mall}
%%   \city{Boston}
%%   \state{MA}
%%   \postcode{02215}
%}
%
%\author{Zichen Zhu}
% \email{zczhu@bu.edu}
%\affiliation{%
%%   \institution{Boston University}
%%   \streetaddress{111 Cummington Mall}
%%   \city{Boston}
%%   \state{MA}
%%   \postcode{02215}
%}

\author{Subhadeep Sarkar, Dimitris Staratzis, Zichen Zhu, Manos Athanassoulis}
\affiliation{%
\vspace*{1mm}
\institution{Boston University, MA, USA}
%  \streetaddress{111 Cummington Mall}
%  \city{Boston}
%  \state{MA, USA}
%  \postcode{02215}
}
\email{ssarkar1@bu.edu, dstara@bu.edu, zczhu@bu.edu, mathan@bu.edu}

\begin{abstract}
  % In this paper, we show that while modern log-structured merge (LSM) tree-based data stores offer a superior write throughput, they pay an exorbitant cost in terms of write amplification. 
% State-fo-the-art production LSM-based data stores exhibit a write amplification up to $42\times$. 
% We identify that, in practice, this write amplification is driven by file picking policy adopted by the data store during compaction, and that modern data stores fail to extract the optimal performance by adhering to a static file picking policy. 

Log-structured merge (LSM) trees offer efficient ingestion by appending incoming data, and thus, are widely used as the storage layer of production NoSQL data stores. 
% LSM-trees offer high write throughput by employing the out-of-place paradigm and by arranging the data on disk in a tree-like structure, having levels with exponentially growing capacity. 
To enable competitive read performance, LSM-trees periodically re-organize data to form a tree with levels of exponentially increasing capacity, through iterative \textit{compactions}.
% Each level is a collection of \textit{immutable sorted runs}, which may only be updated during \textit{compactions}. 
Compactions fundamentally influence the performance of an LSM-engine in terms of write amplification, write throughput, point and range lookup performance, space amplification, and delete performance. 
Hence, choosing the \textit{appropriate compaction strategy} is crucial and, at the same time, hard as the LSM-compaction design space is vast, largely unexplored, and has not been formally defined in the literature. 
As a result, most LSM-based engines use a fixed compaction strategy, typically hand-picked by an engineer, which decides \emph{how} and \emph{when} to compact data.
 
% In this work, we highlight/show that compactions play a defining role in the performance of LSM-based storage engines. Further, the implications of compaction strategies vary with the workload characteristics. 
% and that state-of-the-art LSM-engines fail to extract the optimal performance by adhering to a static compaction strategy. 

In this paper, we present the design space of LSM-compactions, and evaluate state-of-the-art compaction strategies with respect to key performance metrics.
Toward this goal, our first contribution is to introduce a set of four design primitives that can formally define any compaction strategy: (i) the compaction trigger, (ii) the data layout, (iii) the compaction granularity, and (iv) the data movement policy. 
Together, these primitives can synthesize both existing and completely new compaction strategies.
Our second contribution is to experimentally analyze $10$ compaction strategies. 
We present $12$ observations and $7$ high-level takeaway messages, which show how LSM systems can navigate the compaction design space.
\end{abstract}

\maketitle

\pagestyle{\vldbpagestyle}
\begingroup\small\noindent\raggedright\textbf{PVLDB Reference Format:}\\
\vldbauthors. \vldbtitle. PVLDB, \vldbvolume(\vldbissue): \vldbpages, \vldbyear.
\href{https://doi.org/\vldbdoi}{doi:\vldbdoi}
\endgroup
%%%%%%%%%%%%%%%%%%%%%%%%%%%%%%%%%%%%%%%
\begingroup
\renewcommand\thefootnote{}\footnote{\noindent
This work is licensed under the Creative Commons BY-NC-ND 4.0 International License. 
%Visit \url{https://creativecommons.org/licenses/by-nc-nd/4.0/} to view a copy of this license. For any use beyond those covered by this license, obtain permission by emailing \href{mailto:info@vldb.org}{info@vldb.org}. 
Copyright is held by the owner/author(s). Publication rights licensed to the VLDB Endowment. \\
\raggedright Proceedings of the VLDB Endowment, Vol. \vldbvolume, No. \vldbissue\ %
ISSN 2150-8097. \\
\href{https://doi.org/\vldbdoi}{doi:\vldbdoi} \\
}\addtocounter{footnote}{-1}\endgroup

\ifdefempty{\vldbavailabilityurl}{}{
\begingroup\small\noindent\raggedright\textbf{PVLDB Artifact Availability:}\\
The source code, data, and/or other artifacts have been made available at \url{\vldbavailabilityurl}.
\endgroup
}

%\vspace{-0.08in}
\section{Introduction}
%\vspace{-0.05in}
\label{sec:introduction}

\begin{figure}
    \centering
    \includegraphics[scale=0.34]{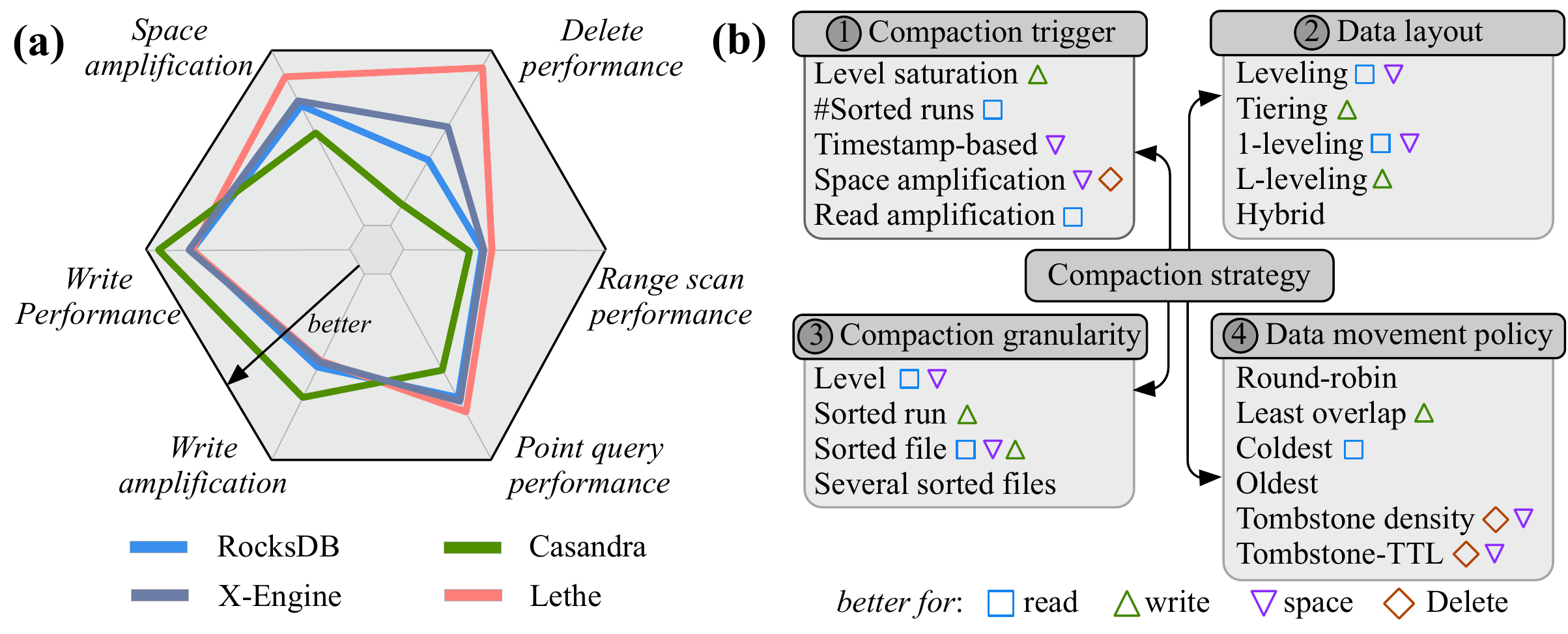}
    \vspace{-0.1in}
    \caption{(a) The different compaction strategies adopted in state-of-the-art LSM-engines lead to the diverse performances offered by the engines; (b) The taxonomy of LSM compactions in terms of the design primitives.}
    \label{fig:intro}
    \vspace{-0.1in}
\end{figure}

\Paragraph{LSM-based Key-Value Stores}
Log-structured merge (LSM) trees are widely used today as the storage layer of modern NoSQL key-value 
stores~\cite{Idreos2020,ONeil1996,Luo2020b}. 
LSM-trees employ the \textit{out-of-place} paradigm to achieve fast ingestion. 
Incoming key-value pairs are buffered in main memory, and are periodically flushed to persistent storage as \textit{sorted immutable runs}. 
As more runs accumulate on disk, they are sort-merged to construct fewer yet longer sorted runs. 
This process is known as \textit{compaction}~\cite{Luo2020b,FacebookRocksDB}. 
To facilitate fast \textit{point lookups}, LSM-trees use auxiliary in-memory data structures (Bloom filters and fence pointers) that help to reduce the average number of disk I/Os performed per lookup~\cite{Dayan2017,Dayan2018a}. 
Because of these advantages, LSM-trees are adopted by several production key-value stores including LevelDB~\cite{GoogleLevelDB} and BigTable~\cite{Chang2006} at Google, RocksDB~\cite{FacebookRocksDB} at Facebook, X-Engine~\cite{Huang2019} at Alibaba, WiredTiger at MongoDB~\cite{WiredTiger}, CockroachDB at Cockroach Labs~\cite{CockroachDB2017}, Voldemort~\cite{LinkedInVoldemort} at LinkedIn, DynamoDB~\cite{DeCandia2007} at Amazon, AsterixDB~\cite{Alsubaiee2014}, Cassandra~\cite{ApacheCassandra}, HBase~\cite{ApacheHBase}, Accumulo~\cite{ApacheAccumulo} at Apache, and bLSM~\cite{Sears2012} and cLSM~\cite{Golan-Gueta2015} at Yahoo. 
Academic systems based on LSM-trees include Monkey~\cite{Dayan2017}, SlimDB~\cite{Ren2017}, Dostoevsky~\cite{Dayan2018, Dayan2018a}, LSM-Bush~\cite{Dayan2019}, Lethe~\cite{Sarkar2020}, Silk~\cite{Balmau2019, Balmau2020}, LSbM-tree~\cite{Teng2017}, SifrDB~\cite{Mei2018}, and Leaper~\cite{Yang2020}.

% \Paragraph{LSM-Basics} For an LSM-tree with $L$ levels, the first level refers to an 
% in-memory buffer with the remaining $L-1$ levels residing on disk.  
% Incoming entries are stored within the memory-resident buffer 
% and are flushed to disk as a \textit{sorted run} every time the buffer gets full.
% To bound the number of levels on disk, the disk-resident levels have 
% exponentially increasing capacity.
% Each level consists of several immutable sorted files. 
% In a typical LSM-tree, all files at the same level 
% form a single sorted run. This ensures that there can be at most one 
% occurrence of a key at any level of the tree. 

\Paragraph{Compactions in LSM-Trees} 
Compactions in LSM-trees are employed periodically to \textit{reduce read and space amplification at the cost of write amplification} while ensuring data consistency and query correctness~\cite{Athanassoulis2016,Athanassoulis2016b}. 
A compaction merges two or more sorted runs, between 
one or multiple levels to ensure that the LSM-tree
maintains levels with exponentially increasing sizes~\cite{ONeil1996}. 
Compactions are typically invoked when a level reaches 
its capacity, at which point, the compaction 
routine moves data from the saturated level to the next one, that has an exponentially larger capacity. 
Any duplicate entries (resulting from \textit{updates}) and invalidated entries (resulting from \textit{deletes}) are removed during a compaction, retaining only the logically correct (latest valid) version~\cite{Dong2017,Sarkar2020}. 
%A compaction either (a) sort-merges two consecutive levels in their entirety -- \textit{full compaction} or (b) sort-merges a part of a level only with the overlapping part from the next level -- \textit{partial compaction}.
Compactions dictate \textit{how} and \textit{when} disk-resident data is re-organized, and thereby, influence the physical data layout on the disk. 
% For example, in state-of-the-art LSM-engines, the amount of data written to disk due to compactions is $10\times$-$40\times$ the original data size. 
% To amortize the data movement due to compaction, few production LSM-engines organize runs into smaller files and perform compactions at the
% granularity of files instead of levels~\cite{Dong2017}. 
%Thus, the implications of compactions on the overall performance of an LSM-engine are deep which, to the best of our knowledge, are largely unexplored in the literature until now. 
Fig. \ref{fig:intro}(a) presents qualitatively the performance implications of the various compaction strategies adopted in state-of-the-art LSM-engines.

\Paragraph{The Challenge: Hand-Picking Compaction Strategies} 
Despite compactions being critical to the performance of LSM-engines, the process of \textit{choosing an appropriate compaction strategy} requires a human in the loop. 
In practice, decisions on \textit{``how to (re-)organize data on disk''}, and thereby, \textit{``which compaction strategies to implement or use''} in a production LSM-based data store are often subject to the expertise of the engineers or the database administrators (DBAs). 
This is largely due to two reasons. 
First, the process of compaction in LSM-trees is often treated as a black-box and is rarely exposed as a tunable knob~\cite{WiredTiger2021}. 
While the LSM-compaction design space is vast, the lack of a formal template for compactions leads to heavily relying on individual expertise, leaving a large part of the design space unexplored. 
Second, there is a lack of analytical and experimental data on how compactions influence the performance of an LSM-engine subject to the underlying design of the storage engine and the workload characteristics. 
Hence, it is difficult, even for experts, to answer design questions such as: 
%\vspace{-1.5mm}
\begin{itemize} [leftmargin=5mm]
    \item[(i)] My LSM-engine is offering lower write performance than expected: \textit{Would a change in the compaction strategy help? If yes, which strategies should be used?}
    \item[(ii)] The workload we used to process has changed: \textit{How does this affect the read throughput of my system? Is there a compaction strategy that can improve the read throughput?} 
    \item[(iii)] We are due to design a new LSM-engine for processing a specific workload: \textit{How should I compact my data for best overall performance? Is there a compaction strategy that I must avoid?} 
\end{itemize}
%\vspace{-1.55mm}
Relying on human expertise to hand-pick the appropriate compaction strategies for each application does not scale, especially for large-scale system deployments.

\Paragraph{Contributions}
To this end, in this work, we formalize the design space of
compactions in LSM-based storage engines. Further, we 
experimentally explore this space, and based on this, 
we present 7 high-level takeaway messages,
and 12 observations that serve as a
comprehensive set of guidelines for 
LSM-compactions, and lay the groundwork for compaction
tuning and automation.

\Paragraphit{\textbf{Conceptual Contribution: Constructing the Compaction Design Space}} 
%The first contribution of this work is the formalization of the compaction process in LSM-trees, and based on that, the construction of the design space for LSM-compactions. 
We identify the 
% different compaction strategies prevalent in state-of-the-art production and academic LSM-based systems~\cite{Callaghan2018}, and analyzing the 
defining characteristics of a compaction, 
% To this end, we identify the design primitives that pertains to the fundamental design-decisions concerning the data layout and the data (re-)organization policies. 
or compaction \textit{primitives}: (i) the \textbf{trigger} (i.e., \textit{when} to compact), (ii) the \textbf{data layout} (i.e., \textit{how} to organize the data after compaction), (iii) the \textbf{granularity} (i.e., \textit{how much} data to compact at a time), and (iv) the \textbf{data movement policy} (i.e., \textit{which} data to compact).
Together, the four primitives define when and how to compact data in an LSM-tree. 
Fig. \ref{fig:intro}(b) presents the taxonomy of LSM-compactions along with the various options for each of the design primitives.
% Together, these four primitives capture any existing and new LSM-compaction strategies. 

% Depending on eagerness of compactions, LSM-trees are classically classified as \textit{leveled} or \textit{tiered}~\cite{Dayan2017}. 
% Furthermore, some production leveled LSM-based data stores 
% amortize the cost for compactions by controlling the amount of data that is moved from the saturated level to the next one, or the \textit{compaction granularity}. 
% A smaller compaction granularity leads to frequent compactions, but compacts fewer bytes per compaction amortizing the I/O cost associated. 
% State-of-the-art LSM-engines typically perform such \textit{partial compactions} at the granularity of a file (typically, having a size between $512$KB and $64$MB)~\cite{Dong2017,Sarkar2020}. 
% The decision on which file to compact, depends on a pre-decided \textit{data movement policy} that varies across systems~\cite{Dong2016}. 

\Paragraphit{\textbf{Experimental Contribution 1: Unifying the Experimental Infrastructure of Multiple Compaction Strategies}} 
To establish a consistent experimental platform,
we integrate several state-of-the-art compaction strategies into a unified 
codebase, based on the widely adopted open-source RocksDB~\cite{FacebookRocksDB}
LSM-engine.
This integration bridges wild variations of implementation and configuration knobs of different compaction strategies across different LSM-engines.
Further, we implement each compaction strategy through the prism of the aforementioned four primitives on top of the same data store to ensure an apples-to-apples comparison.  
%\blue{We choose RocksDB~\cite{FacebookRocksDB} as it (i) has an open-source and well-documented codebase, (ii) has a large and active community, and (iii) is widely adopted in production.} 
We implement \textit{ten state-of-the-art compaction strategies} that are popular among production and academic systems, and are key to the understanding of the LSM-compaction design space. 
We implement these strategies through significant modifications to the latest RocksDB codebase~\cite{FacebookRocksDB}, and expose \textit{more than a hundred design knobs} to enable custom configuration and to ensure a fair evaluation.

\Paragraphit{\textbf{Experimental Contribution 2: Analyzing the Compaction Design Space}} 
We provide a comprehensive experimental analysis of the LSM-compaction design space, which quantifies the impact of each of the design primitives on a number of performance metrics. 
This experimental analysis also serves as a roadmap
for selecting a compaction strategy subject to the workload characteristics and performance goals. 
We perform more than $2000$ experiments with $10$ compaction strategies to take a deep dive on the following.

%\vspace*{-1mm}
\begin{itemize}[leftmargin=*,labelindent=0mm, itemsep=0.2\baselineskip]
    % \item We present a taxonomy for LSM-compactions by identifying the design primitives that defines the process of data organization in an LSM-tree. 
    % \item We implement ten state-of-the-art compaction strategies on top of RocksDB, an open-source and widely-used LSM-engine, to setup a consistent experiment environment for our evaluation. 
    \item \textit{Performance Implications.} 
%    \blue{Compactions play a central role in the overall performance of an LSM-engine. 
    We quantify the impact of compactions on LSM performance in terms of ingestion throughput, query latency, space and write amplification, and delete efficacy in \S\ref{subsec:performance}.
    \item \textit{Workload Influence on Compactions.} 
    While the composition and (ingestion and access) distribution of the workload influence the compaction performance, deciding which compaction strategy to employ is workload-agnostic in existing systems. 
    To analyze the workload's impact on compactions performance, we experiment with a number of representative workloads by varying (i) the size of ingested data, (ii) the proportion of ingestion and lookups, (iii) the proportion of empty and non-empty point lookups, (iv) the selectivity of range queries, (v) the fraction of updates and (vi) deletes, (vii) the key-value size, as well as (viii) the workload distribution (uniform, normal, and Zipfian) in \S\ref{subsec:workload}. 
    \item \textit{Tuning Influence on Compactions.} LSM tuning typically focuses on knobs like memory buffer size, page size, and size ratio which are not believed to be connected with compaction performance. We experiment with these knobs to uncover when compactions are affected (and when not) by these knobs in \S\ref{subsec:tuning}.
    \item \textit{Answering Design Questions.}
	Finally, throughout \S\ref{sec:results} we present various observations and key insights of our experimental evaluation, and in \S\ref{sec:tuning} we discuss a roadmap for designing and choosing compaction in LSM-engines.
%    We put forward a set of key insights from our experimental evaluation that will allow researchers/engineers to make educated decisions about when to use which compaction strategy for enhanced performance.
\end{itemize}

\begin{center}
    \textit{This work defines the LSM compaction design space and presents a thorough account of how the different primitives affect the overall performance of a storage engine.}
\end{center}

% \Paragraph{Suboptimal Compaction Policies} 
% In this paper, we investigate the compaction design space in LSM-trees and show that compactions have critical implications on the overall performance of the storage engine. 
% State-of-the-art production systems typically adhere to a fixed compaction strategy, regardless of the target performance or the workload characteristics. 
% The compaction policy is typically hard-coded into the source code of the storage engine, and in the best case, optimizes for a particular pre-conceived performance goal. 
% The compaction policies in existing systems are workload-agnostic, and these systems are unable to extract optimal performance when subject to different performance goals. 
% \blue{Moreover, changing the compaction strategy, in practice, leads to migrating to a new storage engine in the worst case or amending to the existing codebase at best.} 
% \red{Compactions, i.e., (i) the compaction trigger, (ii) the compaction granularity, (iii) eagerness of compactions, and (iv) the compaction file picking policy, affect the (a) write throughput, (b) point and (c) range lookup performance (i.e., read throughput), (d) space amplification, (e) write amplification, and (f) delete performance of a storage system.}

\Paragraph{Key Takeaways}
Finally, the high-level key takeaways from our analysis are the following.

\Paragraphit{A. There is no perfect compaction strategy} 
When it comes to selecting a compaction strategy for an LSM-engine, there is \textit{no single best}. 
Thus, a compaction strategy needs to be custom-tailored to specific combinations of workload, LSM tuning, and performance goals.

\Paragraphit{B. It is important to look into the compaction ``black-box''}
To understand the performance implications of LSM compactions, it is crucial to ``open the black-box'', and treat them as a set of design primitives. 
Following this approach, we reason about the performance implications of each design primitive independently. 
We identify common pitfalls given a workload and a target performance.

\Paragraphit{C. The right compaction strategy can significantly boost performance}
Switching between compaction strategies as the workload and/or the performance goals shift can boost the performance of an LSM-engine significantly. 
Understanding the behavior and performance implications of the compaction primitives allows for modifications to existing codebases to invoke the appropriate compaction strategy.

% \Paragraphit{D. The LSM compaction design space is vast and full of new strategies} 

% \Paragraph{Our Intuition: } 
% \blue{We can strike a trade-off by optimizing for different performance targets at the cost of write amplification.}
% \red{Here, we talk about the trade-off among WA, SA, WT, RT, and DelPerf.}

% \Paragraph{Online Demo}
% \red{This should work based on the cost model.}

%\vspace{-0.05in}
\section{Background}
%\vspace{-0.02in}
\label{sec:background}

We now present the necessary background of LSM-trees. 
A more detailed survey on LSM-basics can be found in the literature~\cite{Dayan2017,Luo2020b}.

%\vspace{-0.1in}
\Paragraph{LSM-Basics}
To support fast data ingestion, LSM-trees buffer incoming inserts, updates, and deletes (i.e., ingestion, in general) within main memory. 
Once the memory buffer becomes full, the entries contained are sorted on the key and the buffer is flushed as a \textit{sorted run} to the disk-component of the tree. 
In practice, a sorted run is a collection of one or more \textit{immutable files} that have typically the same size.
For an LSM-tree with $L$ levels, we assume that its first level (Level $0$) is an in-memory buffer and the remaining levels (Level $1$ to $L - 1$) are disk-resident~\cite{Dayan2017,Luo2020b}.
On disk, each Level $i$ ($i>1$) has a capacity that is larger than that of Level $i-1$ by a factor of $T$, where $T$ is the size ratio of the tree.

\Paragraph{LSM-Compactions} 
To limit the number of sorted runs on disk (and thereby, to facilitate fast lookups and better space utilization), LSM-trees periodically sort-merge runs (or parts of a run) from a Level $i$ with the overlapping runs from Level $i+1$. 
This process of data re-organization and creating fewer longer sorted runs on disk is known as \textit{compaction}. 
However, the process of sort-merging data requires the data to be moved back and forth between the disk and main memory. 
This results in write amplification, which can be as high as $40\times$ in state-of-the-art LSM-based data stores~\cite{Raju2017}.
% In state-of-the-art LSM-based data stores, the write amplification due to compactions can be as high as $40\times$~\cite{Raju2017}.
% We further expand on this in Section \ref{sec:compaction}.

\begin{figure}
    \centering
    \includegraphics[width=0.97\columnwidth]{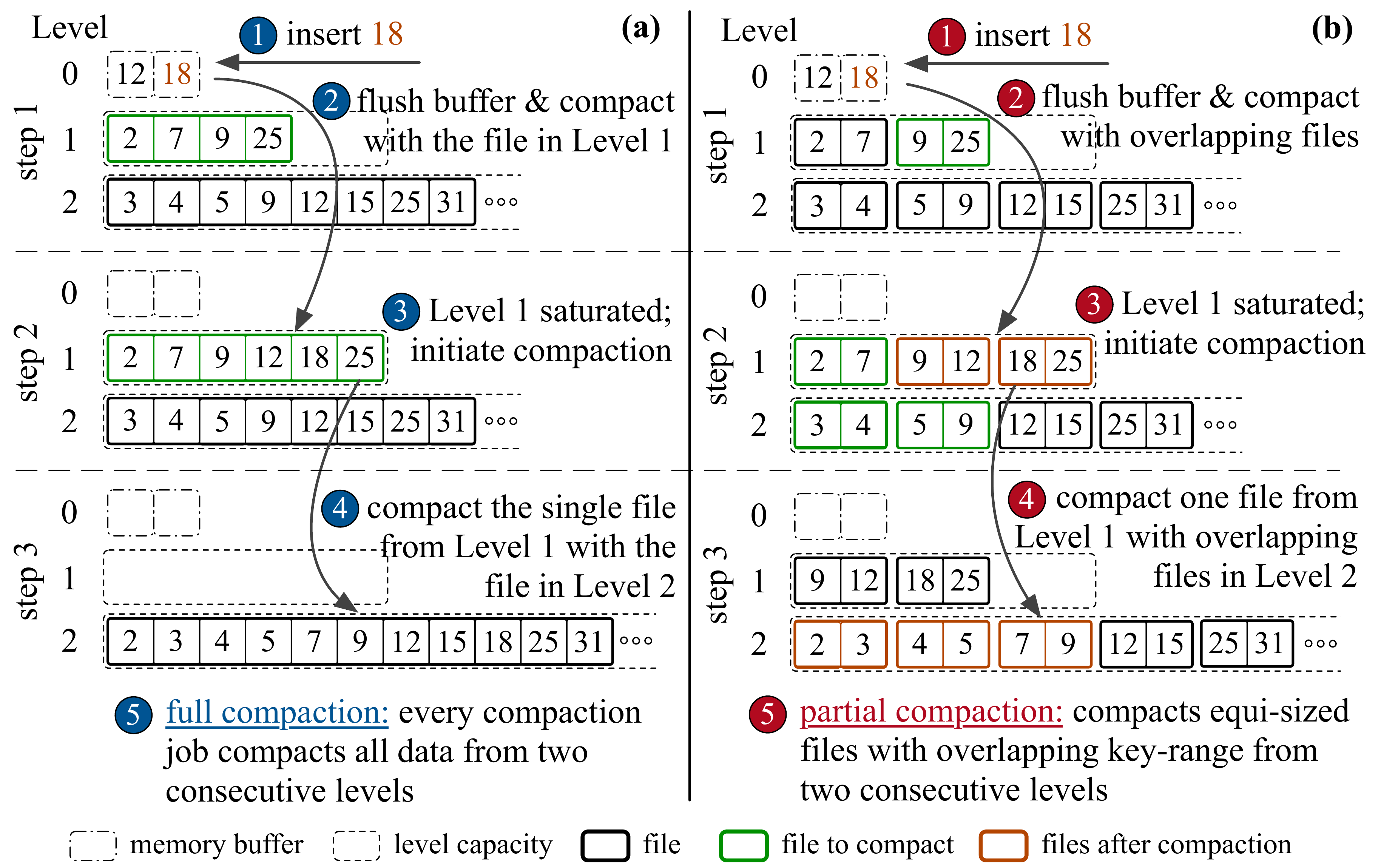}
    \vspace*{-0.1in}
    \caption{(a) When invoked, the classical full compaction routine compacts whole levels at a time, while (b) partial compactions perform compactions at the granularity of files.}
    \label{fig:comp_LSM}
    \vspace*{-0.2in}
\end{figure}

\Paragraphit{Partial compactions}
To amortize data movement, and thus, avoid latency spikes, state-of-the-art LSM-engines organize data into smaller files, and perform compactions at the granularity of files instead of levels~\cite{Dong2017}. 
If Level $i$ grows beyond a threshold, a compaction is triggered and one file (or a subset of files) from Level $i$ is chosen to be compacted with files from
Level $i + 1$ that have an overlapping key-range. 
This process is known as \textit{partial compaction}.
%The decision on which file(s) to compact depends on the design of the storage engine.
Fig.~\ref{fig:comp_LSM} presents a comparative illustration of the full compaction and partial compaction routines in LSM-trees.

\Paragraph{Querying LSM-Trees}
Since LSM-trees realize updates and deletes in an out-of-place manner, multiple entries with the same key may exist in a tree with only the recent-most version being valid. 
%Thus, a query must find and return the recent-most version of an entry. 

\Paragraphit{Point lookups} 
A point lookup starts at the memory buffer and traverses the tree from the smallest level to the largest one, and from the youngest to the oldest run within a level. 
A lookup terminates immediately after a matching key is found. 
To limit the number of runs a lookup probes, state-of-the-art LSM-engines use in-memory data structures, such as Bloom filters and fence pointers~\cite{Dayan2018,FacebookRocksDB}. 

\Paragraphit{Range scans} 
A range scan requires sort-merging the runs qualifying for a range query across all levels of the tree. 
The runs are sort-merged in memory and the latest version for each qualifying entry is returned while discarding all older, logically invalidated versions. 
% To locate the first qualifying disk page for each range query in each run, LSM-trees use fence pointers. 
% Probabilistic data structures that can facilitate both point and range queries, such as SuRF \cite{Zhang2018a} and Rosetta~\cite{Luo2020}, are also employed in LSM-trees. 

\Paragraph{Deletes in LSM-Trees}
%Deletes in LSM-trees are performed logically without necessarily disturbing the target entries. 
A point delete operation is realized by inserting a special type of key-value entry, known as a \textit{tombstone}, that \textit{logically} invalidates the target entries without necessarily disturbing them. 
% A lookup on a deleted key terminates once it finds an entry with a matching key to be marked as a tombstone. 
During compactions, a tombstone purges any older entries with a matching key. 
A delete is eventually considered as \textit{persistent} once the corresponding tombstone reaches the last tree-level, at which point the tombstone can be safely dropped. 
The time taken to persistently delete a data object from an LSM-based data store depends on process of data re-organization.
% Range deletes in LSM-trees are realized by inserting \textit{range tombstones}; however, it requires additional book-keeping in form of special in-memory delete-histograms in order to ensure consistency~\cite{Madan2018}.
Compactions, thus, also play a critical role in timely and persistent deletion of entries, especially in light of the new data privacy regulations~\cite{CCPA2018,Deshpande2018,Kraska2019a,Sarkar2018,Schwarzkopf2019}.

%\vspace{-0.02in}
\section{The Compaction Design Space}
%\vspace{-0.02in}
\label{sec:compaction}

In this section, we identify the \textit{design primitives} that provide a structured decomposition of arbitrary compaction strategies.
This allows us to create the taxonomy of the universe of LSM compaction strategies, including all the classical as well as new ones.

% \vspace{-0.01in}
\subsection{Compaction Primitives}
%\vspace{-0.02in}
We define a compaction strategy as \textit{an ensemble of design primitives that represents the fundamental decisions about the physical data layout and the data (re-)organization policy}. 
Each primitive answers a fundamental design question. 
\begin{itemize}[leftmargin=5mm]
    \item[1)] \textit{Compaction trigger}: \textbf{When} to re-organize the data layout?
    \item[2)] \textit{Data layout}: \textbf{How} to lay out the data physically on storage?
    \item[3)] \textit{Compaction granularity}: \textbf{How much} data to move at-a-time during layout re-organization?
    \item[4)] \textit{Data movement policy}: \textbf{Which} block of data to be moved during re-organization?
\end{itemize} 
Together, these design primitives define \textit{when} and \textit{how} an LSM-engine re-organizes the data layout on the persistent media. 
The proposed primitives capture any state-of-the-art LSM-compaction strategy and also enables synthesizing new or unexplored compaction strategies. 
Below, we define these four design primitives. 
% We consider an $L$-level LSM-tree which has its first level (Level $0$) in memory and the remaining $L-1$ levels (Level $1$ through Level $L-1$) on persistent storage.

\begin{figure*}
    \vspace{-0.2in}
    \centering
    \includegraphics[width=\textwidth]{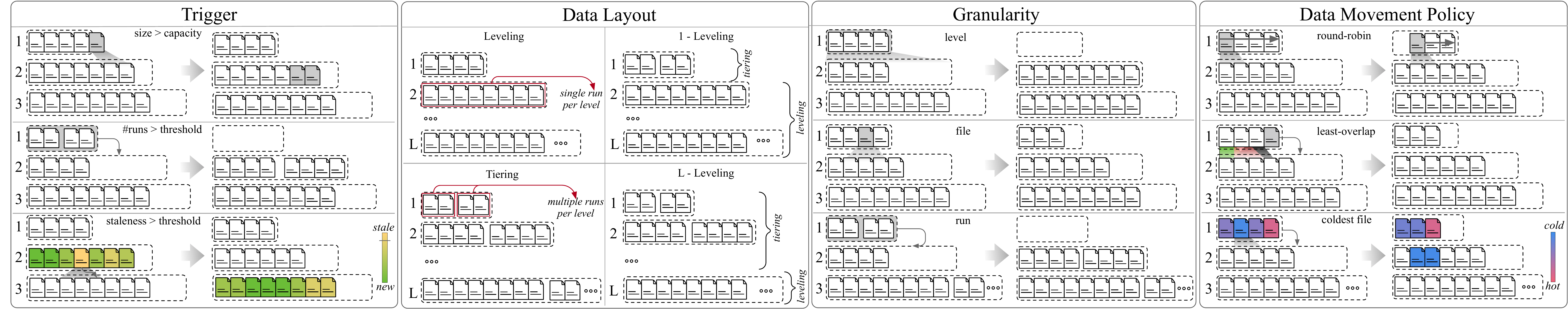}
    \vspace{-0.25in}
    \caption{The primitives that define LSM compactions: trigger, data layout, granularity, and data movement policy.}
    \label{fig:comp_prims}
    \vspace{-0.1in}
\end{figure*}

% \vspace{-0.02in}
\subsubsection{\textbf{Compaction Trigger}} 
Compaction triggers refer to the set of events that can initiate a compaction job. 
The most common compaction trigger is based on the \textit{degree of saturation} of a level in an LSM-tree~\cite{Alsubaiee2014,FacebookRocksDB,GoogleLevelDB,HyperLevelDB,Tarantool,Golan-Gueta2015,Sears2012}. 
The degree of saturation for Level $i$ ($1 \leq i \leq L-1$) is typically measured as the ratio of the number of bytes of data stored in Level $i$ to the theoretical capacity in bytes for Level $i$. 
Once the degree of saturation goes beyond a pre-defined threshold, one or more immutable files from Level $i$ are marked for compaction. 
Some LSM-engines use the file count in a level to compute degree of saturation~\cite{GoogleLevelDB,Huang2019,HyperLevelDB,RocksDB2020,ScyllaDB}. 
Note that the file count-based degree of saturation works only when all immutable files are of equal size, or for systems that have a tunable file size.
The ``\#sorted runs'' compaction trigger, triggers a compaction if the number of sorted runs (or ``tiers'') in a level goes past a predefined threshold, regardless of the size of a level. 
%In practice, a compaction job starts as soon as all the necessary resources, i.e., memory, CPU, and device bandwidth, are available. 

Other compaction triggers include the \textit{staleness of a file}, the \textit{tombstone-based time-to-live}, and \textit{space} and \textit{read amplification}. 
For example, to ensure propagation of updates and deletes to the deeper levels of a tree, some LSM-engines assign a time-to-live (TTL) for each file during its creation. 
Each file can live in a level for a bounded time, and once the TTL expires, the file is marked for compaction~\cite{FacebookRocksDB}. 
Another delete-driven compaction trigger ensures bounded persistence latency of deletes in LSM-trees through a different timestamp-based scheme. 
Each file containing at least one tombstone is assigned a special time-to-live in each level, and up on expiration of this timer, the file is marked for compaction~\cite{Sarkar2020}. 
%Compaction triggers based on space amplification~\cite{RocksDB2020} and read amplification~\cite{Kryczka2020} that can facilitate specific applications have also been implemented in production data stores.
Below, we present a list of the most common \textbf{compaction triggers}:

%\vspace{-0.02in}
\begin{itemize} [leftmargin=5mm]
    \item[i)] \textit{\textbf{Level saturation}}: level size goes beyond a nominal threshold
    \item[ii)] \textit{\textbf{\#Sorted runs}}: sorted run count for a level reaches a threshold
    \item[iii)] \textit{\textbf{File staleness}}: a file lives in a level for too long
    \item[iv)] \textit{\textbf{Space amplification (SA)}}: overall SA surpasses a threshold
    \item[v)] \textit{\textbf{Tombstone-TTL}}: files have expired tombstone-TTL   
\end{itemize}

% \begin{table}[ht]
%     \centering
%     \resizebox{0.475\textwidth}{!}{%
%         % \begin{tabular}{l|cc|cC{0mm}cC{0mm}|cC{0mm}cC{0mm}|cC{0mm}cC{0mm}}
%         \begin{tabular}{l|l|c}
%         \toprule % \Xhline{2\arrayrulewidth} 
%         % HEADER-I
%         \textbf{\textbf{Trigger} }
%         & \multicolumn{1}{c|}{\textbf{Trigger compaction \textit{if}}}
%         & \textbf{Database}      \\  \hline \bottomrule
%         Level saturation
%             & {level size goes beyond a nominal threshold}    & \cite{FacebookRocksDB}    \\ \hline
%         \#Sorted runs
%             & {sorted run count for a level reaches a threshold}    & {}    \\ \hline
%         Staleness of file
%             & {a file lives in a level for too long}    & \cite{FacebookRocksDB}    \\ \hline
%         Space amp.
%             & {space amp. of the database reaches a threshold}    & \cite{FacebookRocksDB}    \\ \hline
%         Write amp.
%             & {avg. write amp. for the database reaches a threshold}    & \cite{FacebookRocksDB}    \\ \hline
%         Read amp.
%             & {expected read amp. reaches a threshold}    & \cite{FacebookRocksDB}    \\ \hline \bottomrule
%         \end{tabular}
%     }
% %    \vspace{-0.1in}
%     \caption{List of common compaction triggers. \label{tab:triggers}}
%     \vspace{-0.25in}
% \end{table}

%\vspace{-0.1in}
\subsubsection{\textbf{Data layout}}
The data layout is driven by the compaction eagerness, and determines the data organization on disk by controlling the number of sorted runs per level.
% LSM-trees employ the out-of-place paradigm for ingestion to the database to achieve a superior write throughput by maximizing the device bandwidth utilization. 
Compactions move data between storage and memory, consuming a significant portion of the device bandwidth. 
There is, thus, an inherent competition for the device bandwidth between ingestion (external) and compaction (internal) -- a trade-off depending on the eagerness of compactions. 

The data layout is commonly classified as \emph{leveling} and 
\emph{tiering}~\cite{Dayan2017,Dayan2018a}. 
With leveling, once a compaction is triggered in Level $i$, the file(s) marked for compaction are merged with the overlapping file(s) from Level $i+1$, and the result is written back to Level $i+1$. 
As a result, Level $i+1$ ends up with a (single) longer sorted run of immutable files~\cite{FacebookRocksDB,Golan-Gueta2015,GoogleLevelDB,Huang2019,HyperLevelDB,Sears2012}. 
For tiering, each level may contain more than one sorted runs with overlapping key domains. 
Once a compaction is triggered in Level $i$, all sorted runs in Level $i$ are merged together and the result is written to Level $i+1$ as a new sorted run without disturbing the existing runs in that level~\cite{Alsubaiee2014,ApacheCassandra,ApacheHBase,FacebookRocksDB,ScyllaDB,Tarantool}. 
A hybrid design is proposed in Dostoevsky~\cite{Dayan2018} where the last level is implemented as leveled and all the remaining levels on disk are tiered. 
%The number of runs in each of these levels is configurable, subject to the workload composition. 
A generalization of this idea is proposed in the literature as a continuum of designs~\cite{Dayan2019,Idreos2019} that allows each level to separately decide between leveling and tiering.
Among production systems, RocksDB implements the first disk-level (Level $1$) as tiering~\cite{RocksDB2020}, and it is allowed to grow perpetually in order to avoid write-stalls~\cite{Balmau2019,Balmau2020,Callaghan2017} in ingestion-heavy workloads. 
%Note that some production LSM-engines internally assign a higher priority to writes over compactions to avoid write stalls; however, this tuning leads to a tree structure that violates the LSM-tree properties, and thus, is out of the scope of our analysis. 
Below is a list of the most common options for \textbf{the data layout}:

%\vspace{-0.02in}
\begin{itemize} [leftmargin=5mm]
    \item[i)] \textit{\textbf{Leveling}}: one sorted run per level
    \item[ii)] \textit{\textbf{Tiering}}: multiple sorted runs per level
    \item[iii)] \textit{\textbf{\bm{$1$}-leveling}}: \textit{tiering} for Level $1$;  \textit{leveling} otherwise
    \item[iv)] \textit{\textbf{\bm{$L$}-leveling}}: \textit{leveling} for last level; \textit{tiering} otherwise
    \item[v)] \textit{\textbf{Hybrid}}: a level can be \textit{tiering} or \textit{leveling} independently
\end{itemize}

\subsubsection{\textbf{Compaction Granularity}}
Compaction granularity refers to the amount of data moved during a single compaction job. 
One way to compact data is by sort-merging and moving all data from a level to the next level -- we refer to this as \textit{full compaction}~\cite{Alkowaileet2020,Alsubaiee2014,Teng2017,WiredTiger}.
This results in periodic bursts of I/Os due to large data movement during compactions, and as a tree grows deeper, the latency spikes are exacerbated causing prolonged write stalls. 
To amortize the I/O costs due to compactions, leveled LSM-based engines employ \textit{partial compaction}~\cite{FacebookRocksDB,GoogleLevelDB,Huang2019,ONeil1996,Sarkar2020,ScyllaDB}, where instead of moving a whole level, a smaller granularity of data participates in every compaction. 
The granularity of data can be a single file~\cite{Dong2017,Huang2019,Sarkar2020} or multiple files~\cite{Alkowaileet2020,Alsubaiee2014,ApacheCassandra,ONeil1996} depending on the system design and the workload. 
Note that, partial compaction does not radically change  the total amount of data movement due to compactions, but amortizes this data movement uniformly over time, thereby preventing undesired latency spikes. 
%The exact data movement due to compaction, is 
%determined by the size ratio of a tree and by the data 
%movement policy we discuss next. 
A compaction granularity of ``sorted runs'' applies principally to LSMs with lazy merging policies. 
Once a compaction is triggered in Level $i$, all sorted runs (or tiers) in Level $i$ are compacted together, and the resulting entries are written to Level $i+1$ as a new immutable sorted run. 
Below, we present a list of the most common \textbf{compaction granularity} options:

%\vspace{-0.02in}
\begin{itemize} [leftmargin=5mm]
    \item[i)] \textit{\textbf{Level}}: all data in two consecutive levels %(thus, no need for a data movement policy)
    \item[ii)] \textit{\textbf{Sorted runs}}: all sorted runs in a level 
    \item[iii)] \textit{\textbf{File}}: one sorted file at a time
    \item[iv)] \textit{\textbf{Multiple files}}: several sorted files at a time
\end{itemize} 

%\vspace{-0.1in}
\subsubsection{\textbf{Data Movement Policy}}
When \textit{partial compaction} is employed, the data movement policy selects 
which file(s) to choose for compaction. 
%Full-level compaction does not need such a policy.
While the literature commonly refers to this decision as \textit{file picking policy}~\cite{Dong2016}, we use the term \textit{data movement} to generalize for any possible data movement granularity. 
%Moving a file from a shallower level (Level $i$) to a deeper level (Level $i+1$) affects point lookup performance, write and space amplification, as well as delete performance. 

A na\"ive way to choose file(s) is at random or by using a round-robin policy~\cite{GoogleLevelDB,HyperLevelDB}. 
These data movement policies do not focus on optimizing for any particular performance metric, but help in reducing space amplification. 
To optimize for read throughput, many production data stores~\cite{Huang2019,FacebookRocksDB} select the ``coldest'' file(s) in a level once a compaction is triggered. 
Another common optimization goal is to minimize write amplification. 
In this policy, files with the least overlap with the target level are marked for compaction~\cite{Callaghan2016,Dong2016}. 
To reduce space amplification, some storage engines choose files with the highest number of tombstones and/or updates~\cite{FacebookRocksDB}.
Another delete-aware approach introduces a tombstone-age driven file picking policy that aims to timely persist logical deletes~\cite{Sarkar2020}. 
%Note that, as LSM-trees store data in immutable files, the smallest granularity of data movement is typically a file. 
%However, with mutable file structures it is possible to move data at a finer granularity, i.e., one or few disk pages at a time. 
Below, we present the list of the common \textbf{data movement policies}:

%\vspace{-0.1in}
\begin{itemize} [leftmargin=5mm]
    \item[i)] \textit{\textbf{Round-robin}}: chooses files in a round-robin manner %space amplfication
    \item[ii)] \textit{\textbf{Least overlapping parent}}: file with least overlap with ``parent''% -- optimizes for write amplification.
    \item[iii)] \textit{\textbf{Least overlapping grandparent}}: as above with ``grandparent'' %-- reduces write amplification. 
    \item[iv)] \textit{\textbf{Coldest}}: the least recently accessed file %-- enhances for point lookup performance.
    \item[v)] \textit{\textbf{Oldest}}: the oldest file in a level %-- reduces space amplification.
    \item[vi)] \textit{\textbf{Tombstone density}}: file with \#tombstones above a threshold %-- reduces space amplification; provides a loose guarantee on timely delete persistence.
    \item[vii)] \textit{\textbf{Tombstone-TTL}}: file with expired tombstones-TTLs% with an expired TTL -- guarantees timely delete persistence; reduces space amplification. 
%    \item[viii)] \textit{\textbf{Entire level}}: when merging two whole consecutive levels
\end{itemize}
%\vspace{-0.15in}

% \vspace{-0.1in}
\subsection{Compaction as an Ensemble of Primitives}
%\vspace{-0.05in}

%\Paragraph{Synthesizing Diverse Compaction Strategies} 
%We now demonstrate the process of harnessing different compaction strategies using these primitives. 
Every compaction strategy takes one or more values for each of the four primitives. 
The trigger, granularity, and data movement policy are multi-valued primitives, whereas data layout is single-valued.

For example, a common LSM design~\cite{Alkowaileet2020} has a \textbf{leveled} LSM-tree (\textit{data layout}) that compacts \textbf{whole levels} at a time (\textit{granularity}) once a \textbf{level reaches a nominal size} (\textit{trigger}). 
This design does not implement many subtle optimizations including partial compactions, and by definition, does not need a data movement policy. 
A more complex example is the compaction strategy for a \textbf{leveled} LSM-tree (\textit{data layout}) in which compactions are performed at the \textit{granularity} of a \textbf{file}. 
A compaction is \textit{triggered} if either (a) a \textbf{level reaches its capacity} or (b) a \textbf{file containing tombstones is retained in a level longer than a pre-set TTL}~\cite{Sarkar2020}.
Once triggered, the \textit{data movement policy} chooses (a) \textbf{the file with the highest density of tombstones}, if there is one or (b) \textbf{the file with the least overlap with the parent level}, otherwise.

\begin{table}[t]
% \vspace{-0.2in}
    \centering
    \resizebox{0.475\textwidth}{!}{%
    \LARGE
        % \begin{tabular}{l|cc|cC{0mm}cC{0mm}|cC{0mm}cC{0mm}|cC{0mm}cC{0mm}}
        \begin{tabular}{l|c|ccccc|cccc|cccccccc}
        \toprule % \Xhline{2\arrayrulewidth} 
        % HEADER-I
        \multicolumn{1}{c|}{\multirow{9}{*}{\begin{tabular}[c]{@{}c@{}}\textbf{Database} \end{tabular}}}    
        & \multirow{9}{*}{\begin{tabular}[c]{@{}c@{}}\textbf{Data layout} \end{tabular}} 
        & \multicolumn{5}{c|}{\textbf{\multirow{1}{*}{\begin{tabular}[c]{@{}c@{}}\textbf{Compaction} \end{tabular}}}}  
        & \multicolumn{4}{c|}{\textbf{\multirow{1}{*}{\begin{tabular}[c]{@{}c@{}}\textbf{Compaction} \end{tabular}}}}  
        & \multicolumn{7}{c}{\textbf{\multirow{1}{*}{\begin{tabular}[c]{@{}c@{}}\textbf{Data Movement} \end{tabular}}}}   \\
        % HEADER-II
        \multicolumn{1}{c|}{} 
        & \multicolumn{1}{c|}{}    
        & \multicolumn{5}{c|}{\textbf{Trigger}}  
        & \multicolumn{4}{c|}{\textbf{Granularity}}  
        & \multicolumn{7}{c}{\textbf{Policy}}  \\   \cline{3-19} 
        % HEADER-III
        &   
        & \rotatebox[origin=l]{90}{Level saturation}    
            & \rotatebox[origin=l]{90}{\#Sorted runs} 
            & \rotatebox[origin=l]{90}{File staleness}    
            & \rotatebox[origin=l]{90}{Space amp.} 
            & \rotatebox[origin=l]{90}{Tombstone-TTL\hspace*{1mm}} 
        & \rotatebox[origin=l]{90}{Level}    
            & \rotatebox[origin=l]{90}{Sorted run} 
            & \rotatebox[origin=l]{90}{File (single)} 
            & \rotatebox[origin=l]{90}{File (multiple)}   
        & \rotatebox[origin=l]{90}{Round-robin} 
            & \rotatebox[origin=l]{90}{Least overlap ($+$$1$)  } 
            & \rotatebox[origin=l]{90}{Least overlap ($+$$2$)} 
            & \rotatebox[origin=l]{90}{Coldest file}    
            & \rotatebox[origin=l]{90}{Oldest file} 
            & \rotatebox[origin=l]{90}{Tombstone density} 
            & \rotatebox[origin=l]{90}{Expired TS-TTL} 
            & \rotatebox[origin=l]{90}{N/A (entire level)} \\    \hline \bottomrule
        % 1.1 RocksDB
        \multirow{2}{*}{RocksDB~\cite{FacebookRocksDB},}     
        & \multirow{1}{*}{Leveling /}        
        & \multirow{2}{*}{\cmark}    & \multirow{2}{*}{}      & \multirow{2}{*}{\cmark}
            & \multirow{2}{*}{}   & {}    & {}  
            & {}    & \multirow{2}{*}{\cmark}   & \multirow{2}{*}{\cmark} & {}
        & \multirow{2}{*}{\cmark}    & \multirow{2}{*}{} & \multirow{2}{*}{\cmark} 
            & \multirow{2}{*}{\cmark} & \multirow{2}{*}{\cmark} & {} \\ 
        % 1.2 Monkey
        \multirow{2}{*}{Monkey~\cite{Dayan2018a}} 
        & \multirow{1}{*}{1-Leveling}  & & & & & & & & & & \\ \cline{2-19}
            
        & \multirow{1.2}{*}{Tiering}    & \multirow{1.2}{*}{}
        & \multirow{1.2}{*}{\cmark} & \multirow{1.2}{*}{}    & \multirow{1.2}{*}{\cmark} & \multirow{1.2}{*}{\cmark} & \multirow{1.2}{*}{} 
        & \multirow{1.2}{*}{\cmark} & \multirow{1.2}{*}{} & \multirow{1.2}{*}{} & \multirow{1.2}{*}{} 
        & \multirow{1.2}{*}{} & \multirow{1.2}{*}{}  & {} & {} & {} & {}  & \multirow{1.2}{*}{\cmark}   \\ \midrule
        % 2.1 LevelDB
        \multirow{1}{*}{LevelDB~\cite{GoogleLevelDB},}     
        & \multirow{2}{*}{Leveling}        
        & \multirow{2}{*}{\cmark}    & \multirow{2}{*}{}      & {}
            & {}   & {}    & {}  
            & {}    & \multirow{2}{*}{\cmark}   & \multirow{2}{*}{} & \multirow{2}{*}{\cmark}
        & \multirow{2}{*}{\cmark}    & \multirow{2}{*}{\cmark} & {} 
            & {} & {} & {} \\ 
        % 2.2 Monkey
        \multirow{1}{*}{Monkey (J.)~\cite{Dayan2017}} 
        & {}  & & & & & & & & & & \\ \midrule
        % 3. SlimDB they do an in-level compaction and then move everything to the next level. The picking policy is not mentioned
        SlimDB~\cite{Ren2017}
        & {Tiering}          
        & \cmark    & {}      & {}
            & {}   & {}    & {}  
            & {}    & \cmark   & \cmark & {}
        & {}    & {} & {} 
            & {} & {} & {} & \cmark \\  \midrule
        % 4. Dostoevsky
        Dostoevsky~\cite{Dayan2018}
        & $L$-leveling    & {\cmark$^{L}$}      & {\cmark$^{T}$}
            & {}   & {}    & {}  
            & {\cmark$^{L}$}    & {\cmark$^{T}$}   & {} & {}
        & {}    & {\cmark$^{L}$} & {} 
            & {} & {} & {} & {} & {\cmark$^{T}$} \\ \midrule 
        % 5. LSM-Bush
        LSM-Bush~\cite{Dayan2019}
        & {Hybrid leveling}    & {\cmark$^{L}$}      & {\cmark$^{T}$}
        & {}   & {}    & {}  
        & {\cmark$^{L}$}    & {\cmark$^{T}$}   & {} & {}
    & {}    & {\cmark$^{L}$} & {} 
        & {} & {} & {} & {} & {\cmark$^{T}$} \\ \midrule
        % 6. Lethe
        Lethe~\cite{Sarkar2020}
        & Leveling    
        & \cmark      & {}
            & {}   & {}    & {\cmark}     &     
        & {}    & \cmark     & \cmark    & {}    
        & {\cmark} &    &     &    
            &     & \cmark \\ \midrule
        % 3. Silk
        Silk~\cite{Balmau2019}, Silk+~\cite{Balmau2020}
        & {Leveling}    
        & {\cmark}      & {}
            & {}   & {}    & {}     & {}    
        & {}    & {\cmark}     & {\cmark}    & {\cmark}    
        & {} & {}   & {}    & {}    
            & {}    & {} \\ \midrule
        % 3. HyperLevelDB
        HyperLevelDB~\cite{HyperLevelDB}
        & {Leveling}    & {\cmark}      & {}
            & {}   & {}    & {}  
            & {}    & {}   & {\cmark} & {}
        & {\cmark}    & {\cmark} & {\cmark} 
            & {} & {} & {} \\ \midrule
        % 3. PebblesDB
        PebblesDB~\cite{Raju2017}
        & {Hybrid leveling}    & {\cmark}      & {}
            & {}   & {}    & {}  
            & {}    & {}   & {\cmark} & {\cmark}
        & {}    & {} & {} 
            & {} & {} & {} & {} & {\cmark} \\ \midrule
        % 4. Cassandra. There is one more compaction trigger which is: The most used file for reads.
        % \multirow{2}{*}{RocksDB~\cite{FacebookRocksDB},} 
        \multirow{2}{*}{Cassandra~\cite{ApacheCassandra}}
        & \multirow{1}{*}{Tiering} 
           & {}      & {\cmark}
            & {\cmark}   & {}    & {\cmark}  
            & {}    & {\cmark}   & {} & {}
        & {}    & {} & {} 
            & {} & {} & {} &{} &{\cmark}
        \\ \cline{2-19}
        
        & \multirow{1.2}{*}{Leveling} 
        & \multirow{1.2}{*}{\cmark}      & \multirow{1.2}{*}{}
            & \multirow{1.2}{*}{}   & \multirow{1.2}{*}{}    & \multirow{1.2}{*}{\cmark}  
            & \multirow{1.2}{*}{}    & \multirow{1.2}{*}{}   & \multirow{1.2}{*}{\cmark} & \multirow{1.2}{*}{\cmark}
        & \multirow{1.2}{*}{}    & \multirow{1.2}{*}{\cmark} & \multirow{1.2}{*}{} 
            & \multirow{1.2}{*}{} & \multirow{1.2}{*}{} & \multirow{1.2}{*}{\cmark} &\multirow{1.2}{*}{\cmark}
              \\ \midrule
        % 5. WiredTiger
        WiredTiger~\cite{WiredTiger}
        & {Leveling}    & {\cmark}      & {}
            & {}   & {}    & {}  
            & {\cmark}    & {}   & {} & {}
        & {}    & {} & {} 
            & {} & {} & {} & {} & {\cmark} \\ \midrule
        % 6. X-Engine
        X-Engine~\cite{Huang2019}, Leaper~\cite{Yang2020}
        & {Hybrid leveling}    & {\cmark}      & {}
            & {}   & {}    & {}  
            & {}    & {}   & {\cmark} & {\cmark}
        & {}    & {\cmark} & {} 
            & {} & {} & {\cmark} \\ \midrule
        % 7. HBase
        HBase~\cite{ApacheHBase}
        & {Tiering}    & {}      & {\cmark}
            & {}   & {}    & {}  
            & {}    & {\cmark}   & {} & {}
        & {}    & {} & {} 
            & {} & {} & {} & {} & {\cmark} \\ \midrule
        % 8. AsterixDB
        \multirow{2}{*}{AsterixDB~\cite{Alsubaiee2014}}
        & \multirow{1}{*}{Leveling}    & {\cmark}      & {{}}
            & {}   & {}    & {}  
            & {\cmark}    & {}   & {} & {}
        & {}    & {} & {} 
            & {} & {} & {} &  {} &{\cmark} \\ \cline{2-19}

        & \multirow{1.2}{*}{Tiering} 
        & \multirow{1.2}{*}{}      & \multirow{1.2}{*}{\cmark}
            & \multirow{1.2}{*}{}   & \multirow{1.2}{*}{}    & \multirow{1.2}{*}{}  
            & \multirow{1.2}{*}{}    & \multirow{1.2}{*}{\cmark}   & \multirow{1.2}{*}{} & \multirow{1.2}{*}{}
        & \multirow{1.2}{*}{}    & \multirow{1.2}{*}{} & \multirow{1.2}{*}{} 
            & \multirow{1.2}{*}{} & \multirow{1.2}{*}{} & \multirow{1.2}{*}{} & \multirow{1.2}{*}{} & \multirow{1.2}{*}{\cmark}
                \\ \midrule

        % 9. Tarantool. about the data movement policy, it says that it picks sorted runs with similar size.
        Tarantool~\cite{Tarantool}
        & {$L$-leveling}    & {\cmark$^L$}      & {\cmark$^T$}
            & {}   & {}    & {}  
            & {\cmark$^L$}    & {\cmark$^T$}   & {} & {}
        & {}    & {} & {} 
            & {} & {} & {} & {} & {\cmark} \\ \midrule 
        % 10. ScyllaDB can dynamically switch between leveling and tiering. see (Temporary fallback to STCS)
        \multirow{2}{*}{ScyllaDB~\cite{ScyllaDB}}
        & \multirow{1}{*}{Tiering}    & {}      & {\cmark}
            & {\cmark}   & {}    & {\cmark}  
            & {}    & {\cmark}   & {} & {}
        & {}    & {} & {} 
            & {} & {} & {} & {} & {\cmark} \\ \cline{2-19} 

        & \multirow{1.2}{*}{Leveling} 
        & \multirow{1.2}{*}{\cmark}      & \multirow{1.2}{*}{}
            & \multirow{1.2}{*}{}   & \multirow{1.2}{*}{}    & \multirow{1.2}{*}{\cmark}  
            & \multirow{1.2}{*}{}    & \multirow{1.2}{*}{}   & \multirow{1.2}{*}{\cmark} & \multirow{1.2}{*}{\cmark}
        & \multirow{1.2}{*}{}    & \multirow{1.2}{*}{\cmark} & \multirow{1.2}{*}{} 
            & \multirow{1.2}{*}{} & \multirow{1.2}{*}{} & \multirow{1.2}{*}{\cmark} & \multirow{1.2}{*}{\cmark} &\multirow{1.2}{*}{} \\ \midrule 

        % 11. bLSM
        bLSM~\cite{Sears2012}, cLSM~\cite{Golan-Gueta2015}
        & {Leveling}    & {\cmark}      & {}
            & {}   & {}    & {}  
            & {}    & {}   & {\cmark} & {}
        & {\cmark}    & {} & {} 
            & {} & {} & {} \\ \midrule
        % % 12. cLSM
        % cLSM~\cite{Golan-Gueta2015}
        % & {Leveling}    & {\cmark}      & {}
        %     & {}   & {}    & {}  
        %     & {}    & {}   & {} & {\cmark}
        % & {\cmark}    & {} & {} 
        %     & {} & {} & {} \\ \midrule
        % % 13. Bigtable
        % Bigtable~\cite{Chang2006}
        % & {Tiering}    & {}      & {}
        %     & {}   & {}    & {}  
        %     & {}    & {}   & {} & {}
        % & {}    & {} & {} 
        %     & {} & {} & {} \\ \midrule
        % 14. Accumulo
        Accumulo~\cite{ApacheAccumulo}
        & {Tiering}    & {\cmark}      & {\cmark}
            & {}   & {}    & {\cmark}  
            & {}    & {\cmark}   & {} & {}
        & {}    & {} & {} 
            & {} & {} & {} & {} & {\cmark}  \\ \midrule
        % 15. LSbM-tree
        LSbM-tree~\cite{Teng2017,Teng2018}
        & {Leveling}    & {\cmark}      & {}
            & {}   & {}    & {}  
            & {\cmark}    & {}   & {} & {}
        & {}    & {} & {} 
            & {} & {} & {} & {} & {\cmark} \\ \midrule
        % 16. SifrDB
        SifrDB~\cite{Mei2018}
        & {Tiering}    & {\cmark}      & {}
            & {}   & {}    & {}  
            & {}    & {}   & {} & {\cmark}
        & {}    & {} & {} 
            & {} & {} & {} & {}  & {\cmark}  \\ \hline
        \bottomrule
        \end{tabular}
    }
%    \vspace{0.05in}
    \caption{Compaction strategies in state-of-the-art systems. [\footnotesize{\cmark$^{L}$\normalsize: for levels with leveling; \footnotesize\cmark$^{T}$\normalsize: for levels with tiering.}\normalsize] \label{tab:db}}
    \vspace{-0.35in}
\end{table}

\Paragraph{The Compaction Design Space Cardinality} 
% The compaction primitives bears the answers to the fundamental design questions for LSM-compactions, and enables us to express any compaction strategy as an ensemble of these primitives. 
Two compaction strategies are considered different from each other if they differ in at least one of the four primitives. 
Compaction strategies that 
%are identical in terms of three of the four primitives, but 
differ in only one primitive, can have vastly different performance when subject to the same workload while running on identical hardware.
Plugging in some typical values for the cardinality of the primitives, we estimate the cardinality of the compaction universe as >$10^4$, a vast yet largely unexplored design space. 
% i.e., we can generate more than $10^4$ different LSM-compaction strategies simply by plugging in different values to the compaction primitives, one at a time. 
Table \ref{tab:db} shows a representative part of this space, detailing the compaction strategies used in more than twenty academic and production systems.

% The primary advantage that we get by breaking the design space in terms of the primitives is that it allows us to formulate not only the classical compaction algos that exists today, but also new compaction algos that are not yet explored but may be useful for certain workloads. 

\Paragraph{Compactions Analyzed} 
For our analysis and experimentation, we select ten representative compaction strategies that are prevalent in production and academic LSM-based systems.
%These strategies capture the wide variety of the possible compaction designs. 
% \red{Assessing the compactions used in Literature and production systems and that have characteristics features pedagogical enhancement}
% that are frequent in research/production systems -- to capture the wide variety of the design choices including the initial compactions proposed ...
% We believe future exploration of the design space may generate equally interesting \dots
We codify and present these candidate compaction strategies in Table \ref{tab:comp_list}. 
\texttt{Full} represents the compaction strategy for leveled LSM-trees that compacts entire levels upon invocation. 
\texttt{LO+1} and \texttt{LO+2} denote two partial compaction routines that choose a file for compaction with the smallest overlap with files in the parent ($i+1$) and grandparent ($i+2$) levels, respectively. 
\texttt{RR} chooses files for compaction in a round-robin fashion from each level. 
\texttt{Cold} and \texttt{Old} are read-friendly strategies that mark the coldest and oldest file(s) in a level for compaction, respectively. 
\texttt{TSD} and \texttt{TSA} are delete-driven compaction strategies with triggers and data movement policies that are determined by the density of tombstones and the age of the oldest tombstone contained in a file, respectively. 
\texttt{Tier} represents a variant of tiered data layout, where compactions are triggered when either (a) the number of sorted runs in a level or (b) the estimated space amplification in the tree reaches certain thresholds. 
This interpretation of tiering is also referred to as \textit{universal compaction} in systems like RocksDB~\cite{Kryczka2020,RocksDB2020}. 
Finally, \texttt{1-Lvl} represents a hybrid data layout where the first disk level is realized as \textit{tiered} while the others as \textit{leveled}. 
This is the default data layout for RocksDB~\cite{Kryczka2020,RocksDB2020a}.

%\vspace{-0.2in}
\begin{table*}[!ht] 
    \centering
    \resizebox{\textwidth}{!}{%
        \begin{tabular}{L{2cm}|C{1.9cm}|C{1.6cm}|C{1.3cm}|C{1.2cm}|C{2.8cm}|C{1.5cm}|C{1.7cm}|C{2.8cm}|C{1.8cm}|C{2.9cm}}
        \toprule
        \multirow{2}{*}{\textbf{Primitives}}   & \textbf{\texttt{Full} \cite{Alsubaiee2014,Teng2017,WiredTiger}}    & \textbf{\texttt{LO+1} \cite{FacebookRocksDB,Dayan2018a,Sarkar2020}}  & \textbf{\texttt{Cold} \cite{FacebookRocksDB}}    & \textbf{\texttt{Old} \cite{FacebookRocksDB}}    & \textbf{\texttt{TSD} \cite{FacebookRocksDB,Huang2019}}    & \textbf{\texttt{RR} \cite{GoogleLevelDB,HyperLevelDB,Sears2012,Golan-Gueta2015}}    & \textbf{\texttt{LO+2} \cite{GoogleLevelDB,HyperLevelDB}}    & \textbf{\texttt{TSA} \cite{Sarkar2020}}   & \textbf{\texttt{Tier} \cite{Ren2017,ApacheCassandra,HBase2013}}    & \textbf{\texttt{1-Lvl} \cite{FacebookRocksDB,Kryczka2020,RocksDB2020a}}   \\ 
        \midrule
        \multirow{2}{*}{Trigger}    & \multirow{2}{*}{level saturation}   & \multirow{2}{*}{level sat.}    & \multirow{2}{*}{level sat.}   & \multirow{2}{*}{level sat.}   & \multirow{1}{*}{1. TS-density}    & \multirow{2}{*}{level sat.}   & \multirow{2}{*}{level sat.}   & \multirow{1}{*}{1. TS age}   & \multirow{1}{*}{1. \#sorted runs}  & \multirow{1}{*}{1. \#sorted runs$^T$}   \\  
        & & & & & 2. level sat. & & & 2. level sat. & \multirow{1}{*}{2. space amp.}  & \multirow{1}{*}{2. level sat.$^L$}   \\
        \midrule
        \multirow{1}{*}{Data layout}    & leveling   & leveling    & leveling   & leveling   & leveling    & leveling   & leveling   & leveling   & tiering & hybrid  \\ 
        \midrule
        \multirow{2}{*}{Granularity}    & \multirow{2}{*}{levels}   & \multirow{2}{*}{files}   & \multirow{2}{*}{files}   & \multirow{2}{*}{files}   &  \multirow{2}{*}{files}   &  \multirow{2}{*}{files}  &  \multirow{2}{*}{files}  &  \multirow{2}{*}{files}  &  \multirow{2}{*}{sorted runs} & \multirow{1}{*}{1. sorted runs$^T$} \\ 
        & & & & & & & & & & \multirow{1}{*}{2. files$^L$}   \\
        \midrule
        \multirow{1}{*}{Data~movement}    & \multirow{2}{*}{N/A}   & \multirow{1}{*}{least overlap.}   & \multirow{2}{*}{coldest file}   &  \multirow{2}{*}{oldest file}  & \multirow{1}{*}{1. most tombstones}    & \multirow{2}{*}{round-robin}   & \multirow{1}{*}{least overlap.}   & \multirow{1}{*}{1. expired TS-TTL}   & \multirow{2}{*}{N/A}  & \multirow{1}{*}{1. N/A$^T$} \\ 
        \multirow{1}{*}{policy}	& & \multirow{1}{*}{parent} & & & \multirow{1}{*}{2.~least~overlap.~parent} & & \multirow{1}{*}{grandparent}   & \multirow{1}{*}{2.~least~overlap.~parent}    & & \multirow{1}{*}{2. least overlap. parent$^L$}\\
        \bottomrule
        \end{tabular}
    }    
%    \vspace{0.05in}
    \caption{Compaction strategies evaluated in this work. [\footnotesize{$^{L}$\normalsize: levels with leveling; \footnotesize$^{T}$\normalsize: levels with tiering.}\normalsize]\label{tab:comp_list}}
    \vspace{-0.25in}
\end{table*}

\section{Benchmarking Compactions}
%\vspace{-0.02in}
\label{sec:methodology}

We now discuss our experimental 
platform, how we integrated new compactions policies,
and our measurement methodology.

% \vspace{-0.1in}
%\subsection{Compaction Benchmark}
\subsection{Standardization of Compaction Strategies}
% This section outlines our rationale for choosing RocksDB as the unifying platform for our experiments and the details of the implementation process.
%\vspace{-0.05in}

%\Paragraph{Standardization of Compaction Strategies} 
We choose RocksDB \cite{FacebookRocksDB} as our experimental platform, as it (i) is open-source, (ii) is widely used across industry and academia, (iii) has a large active community. 
To ensure fair comparison we implement all compaction
strategies under the same LSM-engine.

%It is important to implement the different compaction strategies on a single LSM-engine (as opposed to compare their performance when implemented on different engines) to ensure accurate and fair comparison, because the design and implementation of different LSM-operations vary vastly across different engines. 
%Moreover, setting hundreds of tunable knobs to the same set of values for different LSM-engines is virtually impossible. 
%Thus, evaluating the selecting compaction strategies using the same platform allows for an apples-to-apples comparison. 

\Paragraph{Implementation} 
We integrate our codebase into RocksDB v6.11.4. 
We assign to \textit{compactions a higher priority than writes} 
to accurately benchmark them, while always maintaining the 
LSM structure~\cite{Sarkar2021}. 

%measure their performance implications, while guaranteeing that the LSM structure is maintained during workload execution~\cite{Sarkar2021}.

\Paragraphit{Compaction Trigger} 
The default compaction trigger for (hybrid) leveling in RocksDB is level saturation~\cite{RocksDB2020a}, and for the universal compaction is space amplification~\cite{RocksDB2020}. 
RocksDB also supports delete-driven compaction triggers, specifically whether the \#tombstones in a file goes beyond a threshold. 
We further implement a trigger based on the tombstones age to facilitate timely deletes~\cite{Sarkar2020}. 
% For this, we used the codebase developed by the authors of \cite{Sarkar2020}, which required changes to the file metadata structure of RocksDB. 

\Paragraphit{Data layout} 
By default, RocksDB supports only two different data layouts: \textit{hybrid leveling} (tiered first level, leveled otherwise)~\cite{RocksDB2020a} and a variation of \textit{tiering} (with a different trigger), termed \emph{universal compaction}~\cite{RocksDB2020}. 
We also implement pure \textit{leveling} by limiting the number of first-level runs to one, and triggering a compaction when the number of first-level files is more than one. 

\Paragraphit{Compaction Granularity}
The granularity for leveling is \textit{file} and \textit{sorted runs} for tiering.
To implement classical leveling, we mark all files of a level for compaction. 
We ensure that ingestion may resume only after all the compaction-marked files are compacted thereby replicating the behavior of the full compaction routine. 

\Paragraphit{Data Movement Policy}
RocksDB (v6.11.4) provides four different data movement policies: a file (i) with   least overlap with its parent level, (ii) least recently accessed, (iii) with the oldest data in a level, and (iv) that has more tombstones than a threshold. 
We also implement partial compaction strategies that choose a file (v) in a round-robin manner, (vi) with the least overlap with its grandparent level, and (vii) based on the age of the tombstones in a file.

\Paragraph{Designing the Compaction API} 
We expose the compaction primitives through a new API
as configurable knobs. 
An application can configure the desired compaction strategy and initiate workload execution. 
The API also allows the application to change the compaction strategy for an existing database. 
%Further, we expose more than a hundred design knobs for RocksDB through this API. 
Overall, our experimental infrastructure allows us (i) to ensure an identical underlying structure while setting the compaction benchmark, and (ii) to tune and configure the design of the LSM-engine as necessary.

% \vspace{-0.1in}
\subsection{Performance Metrics}
%\vspace{-0.04in}
%Compactions affect both LSM reads and writes.
%Despite being directly related to ingestion, compactions affect both write and read performance of an LSM-engine along with other performance metrics, such as space amplification and delete performance.
We now present the performance metrics used in our analysis.

\Paragraph{Compaction Latency} 
The compaction latency includes the time taken to (i) identify the files to compact, (ii) read the participating files to memory, (iii) sort-merge (and remove duplicates from) the files, (iv) write back the result to disk as new files, (v) invalidate the older files, and (vi) update the metadata in the manifest file~\cite{FacebookRocksDB}.
%, and (vi) some more bookkeeping. 
\textit{The RocksDB metric \texttt{rocksdb.compaction.times.micros} is used to measure the compaction latency.}

% \Paragraph{Write Metrics} 
% The data movement due to compaction causes high write amplification, which results in under-utilization of the device bandwidth and leads to poor write throughput~\cite{Raju2017}. 

\Paragraph{Write Amplification (WA)} 
%A compaction job takes as input a number of immutable files that are read (or streamed) into memory from disk, and writes back a new set of immutable files to disk after the compaction job is completed. 
The repeated reads and writes due to compaction cause high WA~\cite{Raju2017}. 
We formally define WA as \textit{the number of times an entry is (re-)written without any modifications to disk during its lifetime}. 
\textit{We use \texttt{rocksdb.compact.write.bytes} and the actual data size to compute WA.}

\Paragraph{Write Latency} 
Write latency is driven by the device bandwidth utilization, which depends on (i) write stalls due to compactions and (ii) the sustained device bandwidth. 
% While the time taken to complete a compaction job influences the frequency and amplitude of the write stalls, the overall data movement due to compaction determines the device bandwidth that can be used for writes to the database. 
\textit{We use the \texttt{rocksdb.db.write.micros} histogram to measure the average and tail of the write latency.} 
%\green{for Manos: Do you think we need to talk about tail write/compaction latency here?}

\Paragraph{Read Amplification (RA)} 
RA is the ratio between the total number of disk pages read for point lookups and the pages that should be read \emph{ideally}.
\textit{We use \texttt{rocksdb.bytes.read} to compute RA.}

\Paragraph{Point Lookup Latency}
Compactions determine the position of the files in an LSM-tree which affects point lookups on entries contained in those files. 
%In practice, a point lookup may fetch the (Bloom) filter blocks and index (fence pointer) blocks before fetching the data page -- which significantly adds to the I/O latency for point lookups. 
% \Paragraphit{Definition} 
% The point lookup latency is defined as the \textit{time taken to fetch the necessary disk pages from disk to memory, and return the result for the lookup}. 
\textit{Here, we use the \texttt{rocksdb.db.get.micros}.}

\Paragraph{Range Lookup Latency}
The range lookup latency depends on the selectivity of the range query, but is affected by the data layout.
%and the number of sorted runs in a tree. 
%The compaction eagerness determines the number of sorted runs in a tree, and thereby, affects the range lookup performance -- especially for queries with a smaller selectivity. 
\textit{We also use the \texttt{db.get.micros} histogram for range lookups.}

\Paragraph{Space Amplification (SA)}
%As an out-of-place structure, in presence of updates and deletes, an LSM-tree may host several logically invalidated entries leading to under-utilization of the disk space. 
SA depends on the data layout, compaction granularity, and the data movement policy. 
SA is defined as \textit{the ratio between the size of logically invalidated entries and the size of the unique entries in the tree}~\cite{Dayan2018}.
\textit{We compute SA using the size of the database  and the size of the logically valid entries.}

\Paragraph{Delete Performance}
We measure the degree to which the tested compaction strategies 
persistently delete entries within a time-limit~\cite{Sarkar2020} in order to analyze the implications of compactions from a privacy standpoint~\cite{CCPA2018,Deshpande2018,Kraska2019a,Sarkar2018,Schwarzkopf2019,Wang2019}. 
%To quantify the delete performance, we use the metric \textit{delete persistence latency}as the time elapsed between insertion of a tombstone and when all entries with a matching
%(older) key are physically deleted from the database~\cite{Sarkar2020}. 
% To ensure timely and persistent deletion, a tombstone must participate in a compaction involving the last tree-level within the threshold time.
\textit{We use the RocksDB file metadata \texttt{age} and a delete persistence threshold.}

% \vspace{-0.1in}
\subsection{Benchmarking Methodology}
We now discuss the methodology for varying the
key input parameters for our analysis: \textit{workload} and the \textit{LSM tuning}.

%\vspace{-0.02in}
\subsubsection{\textbf{Workload}}
A typical key-value workload comprises of five primary operations: inserts, updates, point lookups, range lookups, and deletes. 
Point lookups target keys that may or may not exist in the database -- we refer to these as \textit{non-empty} and \textit{empty point lookups}, respectively.
Range lookups are characterized by their \textit{selectivity}.
%, and deletes target keys that are present in the database. 
To analyze the impact of each operation, we vary the \textit{fraction} of each operation as well as their qualitative characteristics (i.e., selectivity and entry size). 
We further vary the \textit{data distribution} of ingestion and queries focusing on (i) uniform, (ii) normal, and (iii) Zipfian distributions.
Overall, our custom-built benchmarking suite is a superset of the influential YCSB benchmark~\cite{Cooper2010} as well as the insert benchmark~\cite{Callaghan2017a}, and supports a number of parameters that are missing from existing workload generators, including deletes. 
Our workload generator exposes over $64$ degrees of freedom, and is available via GitHub~\cite{Sarkar2021a} for dissemination, testing, and adoption.

%\vspace{-0.02in}
\subsubsection{\textbf{LSM Tuning}}
We further study the interplay of LSM tuning and compaction strategies.
We consider questions like
\textit{which compaction strategy is appropriate for a specific LSM design and a given workload?}
To answer such questions we vary in our
experimentation key LSM tuning parameters, like (i) the memory buffer size, (ii) the block cache size, and (iii) the size ratio of the tree.

\section{Experimental Evaluation}
%\vspace{-0.02in}
\label{sec:results}

We now present the key experimental results using the ten compaction strategies listed in Table \ref{tab:comp_list}. 

\Paragraph{Goal of the Study} 
% \green{Should this para be moved to the beginning of Section 4?}
Our analysis aims to answer the following three fundamental questions:
\begin{itemize}
    \item[i)] \textit{\underline{\smash{Performance implications}}}: How do compactions affect the overall performance of LSM-engines?
    \item[ii)] \textit{\underline{\smash{Workload influence}}}: How do workload distribution and composition influence compactions, and thereby, the performance of LSM-engines?
    \item[iii)] \textit{\underline{\smash{Tuning influence}}}: What is the interplay between LSM compactions and tuning?
\end{itemize}
%\vspace{-0.05in}
Ultimately, the goal of this study is to help practitioners and 
researchers to make informed
decisions when deciding which compaction strategies to support and 
use in an LSM-based engine.

%We use the word \textit{appropriate}, instead of \textit{optimal} or \textit{best}, because our goal is to make educated decisions to \textbf{avoid the worst choices} as opposed to \textit{finding the optimal/best choice}. 
%Finding the optimal compaction strategy would require \textit{a priori} knowledge of the exact (composition and distribution of the) workload along with a complete knowledge of the underlying hardware and engine design specifications. 
%These requirements are almost never met in practice. 
%Thus, the aim of this paper is to allow the engineers and researchers to make educated choices empowered by the insights and key takeaways presented through this section.

\Paragraph{Experimental Setup} 
For our experiments, we use an AWS EC2 server with t2.2xlarge instances (virtualization: hardware virtual machine)~\cite{EC2a}. 
Each virtual machine has $8$ Intel Scalable Processors (vCPUs) at $3.0$GHz, 
$32$GB of DIMM RAM, $45$MB of L3 cache, and runs Ubuntu 20.04 LTS. 
For storage, we attach a $40$GB SSD volume with $4000$ provisioned IOPS (volume type: io2)~\cite{EC2}. 
For experiments with data size larger than $16$GB, we switch to a $500$GB SSD.

\Paragraph{Default Setup} 
Unless otherwise mentioned, all experiments are performed on a RocksDB setup with an LSM-tree of size ratio $10$~\cite{Dayan2017,Dong2016,FacebookRocksDB}. 
The memory buffer is implemented as a hash skiplist~\cite{Facebook2021}. 
The size of the write buffer is set to $8$MB which can hold up to $512$ $16$KB disk pages~\cite{Dayan2017,Dayan2018a,Dong2016,Huang2019}. 
Fence pointers are maintained for each disk page, and Bloom filters are constructed for every file with $10$ bits memory allocated for every entry~\cite{Dayan2018a,Dayan2019}. 
Additionally, we have $8$MB block cache (RocksDB default) assigned for data, filter, and index blocks~\cite{Dong2016}. 
To capture the true raw performance of RocksDB as an LSM-engine, we (i) assign compactions a higher priority than writes, (ii) enable direct I/Os for both read and write operations, (iii) limit the number of memory buffers (or memtables) to two (one immutable and one mutable), and (iv) set the number of background threads responsible for compactions to $1$.

\Paragraph{Workloads}
%We experiment with variations of YCSB.
%We develop experiment with variations of YCSB key-value benchmark \cite{Cooper2010}. 
Unless otherwise mentioned, ingestion and lookups are uniformly generated, and
the average size of a key-value entry is $128$B with $4$B keys~\cite{Dai2020,Lu2016,Luo2020}. 
We vary the number of inserts, going up to $2^{28}$.  
As compaction performance proves to be agnostic to data size, 
and in the interest of experimenting with many configurations,
we perform our base experiments with $10$M inserts~\cite{Cao2020,Dayan2017}, both interleaved and serial with respect to lookups. 
%We experiment with both interleaved and serial execution of ingestion and lookups.
%Unless otherwise mentioned, ingestion and lookups are uniform and interleaved. 
Further specifications of the workloads are presented before each set of experiments. 

\Paragraph{Presentation} 
For each experiment, we present the primary observations (\textbf{O}) along with key takeaway (\textbf{TA}) messages. 
In the interest of space, we limit our discussion to the most interesting results. 
%More results are available in our technical report~\cite{Sarkar2021b}.
Further, note that \texttt{TSD} and \texttt{TSA}, fall back to \texttt{LO+1} in absence of deletes, and thus, are omitted from the experiments without deletes. 

%\vspace{-0.05in}
\subsection{Performance Implications}
%\vspace{-0.03in}
\label{subsec:performance}
We first analyze the implications of compactions on the ingestion, lookup, and overall performance of an LSM-engine. 

%\Paragraph{Setup} 
%Default.

% \vspace{-0.05in}
\subsubsection{\textbf{Data loading}}
In this experiment, we insert $10$M key-value entries uniformly generated into an empty database to quantify the raw ingestion and compaction performance. 
%The influence of the different compaction policies on ingestion is shown in Fig.~\ref{fig:W1}(a)-(d). 

\begin{figure*}[!ht]
\vspace{-0.3in}
    \centering
    \includegraphics[width=0.98\textwidth]{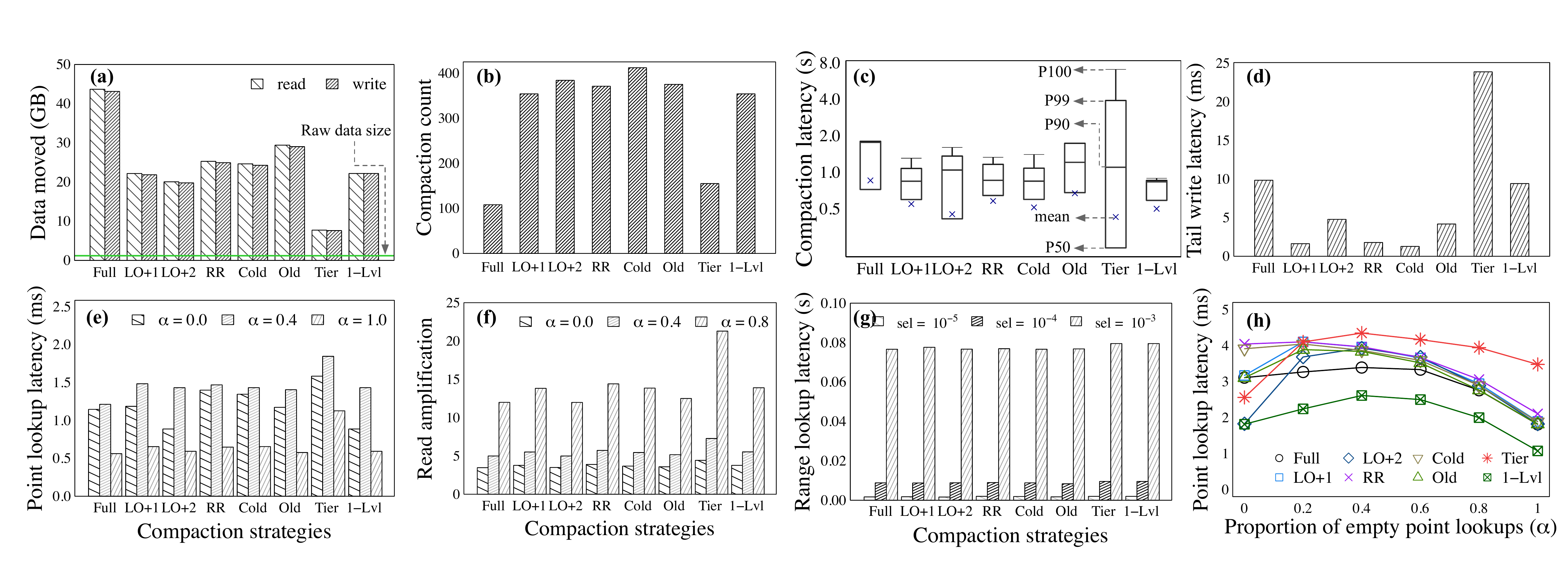}
    % \caption{bytes in compaction}
    \label{fig:W1-bytes-comp}
    \vspace*{-0.25in}
    \caption{Compactions influence the ingestion performance of LSM-engines heavily in terms of (a) the overall data movement, (b) the compaction count, (c) the compaction latency, and (d) the tail latency for writes, as well as (e, f) the point lookup performance. The range scan performance (g) remains independent of compactions as the amount of data read remains the same. Finally, the lookup latency (h) depends on the proportion of empty queries ($\alpha$) in the workload.}
    \label{fig:W1}
    \vspace{-0.1in}
\end{figure*}

\Paragraph{\Ob Compactions Cause High Data Movement} 
Fig.~\ref{fig:W1}(a) shows that the overall (read and write) data 
movement due to compactions is significantly larger than the actual 
size of the data ingested. 
Among the leveled LSM-designs, \texttt{Full} moves $63\times$ ($32\times$ for reads and $31\times$ for writes) the data originally ingested.
The data movement is significantly smaller for \texttt{Tier}, however, it remains $23\times$ of the data size.
The data movement for \texttt{1-Lvl} is similar to that of the leveled strategies in partial compaction.
These observations conforms with prior work~\cite{Raju2017}, but also highlight the problem of \textit{read amplification due to compactions} leading to poor device bandwidth utilization.

% \vspace{0.5mm}
% \noindent\fbox{%
%     \parbox{0.465\textwidth}{%
%     \small
%         \Paragraph{\TA \textit{Compactions with higher eagerness move more data}} \textit{Eager compactions (leveling) compact $2.5$--$5.5\times$ more data than lazier ones (tiering).} 
%     }%
%     \normalsize
% }
% \vspace{0.3mm}

\Paragraph{\Ob Partial Compaction Reduces Data Movement at the Expense of Increased Compaction Count} 
We now shift our attention to the different variations
of leveling. 
%Contrary to \texttt{Full}, all other approaches do not compact 
%entire levels but only a small number of overlapping files.
Fig. \ref{fig:W1}(a) shows that leveled partial compaction leads to $34\%$--$56\%$ less data movement than \texttt{Full}. 
The reason is twofold:
(1) A file with no overlap with its parent level, is only logically merged. Such \textit{pseudo-compactions} require simple metadata (file pointer) manipulation in memory, and no I/Os. 
(2) A smaller compaction granularity reducing overall data movement by choosing a file with (i) the least overlap, (ii) the most updates, or (iii) the most tombstones for compaction.
Specifically, \texttt{LO+1} (and \texttt{LO+2}) is designed to pick files with the least overlap with the parent $i+1$ (and grandparent $i+2$) level. 
They move $10\%$--$23\%$ less data than other partial compaction strategies. 

% \vspace{1mm}
% \noindent\fbox{%
%     \parbox{0.465\textwidth}{%
%     \small
%         \Paragraph{\TA \textit{Smaller compaction granularity reduces data movement}} \textit{Partial compactions move $\sim$$42\%$ less data than full-level compactions.} 
%     }%
%     \normalsize
% }
% \vspace{1mm}

% \Paragraph{\Ob The Compaction Count is Higher for Partial Compaction Routines} 
Fig. \ref{fig:W1}(b) shows that the partial compaction strategies as well as \texttt{1-Lvl} perform $4\times$ more 
compaction jobs than \texttt{Full}, which is equal to the number of tree-levels. 
Note that for an LSM-tree with partial compaction, every
buffer flush triggers cascading compactions to all $L$ levels, while in a 
full-level compaction system this happens when a level is full (every $T$
compactions). 
Finally, since both \texttt{Tier} and \texttt{Full} are full-level
compactions the compaction count is similar.

\vspace{1mm}
\noindent\fbox{%
    \parbox{0.465\textwidth}{%
    \small
        \Paragraph{\TA \textit{Larger compaction granularity leads to fewer but larger compactions}} \textit{Full-level compactions perform about $1/L$ times fewer compactions than partial compaction routines, however, full-level compaction moves nearly $2L$ times more data per compaction.} 
    }%
    \normalsize
}
\vspace{0.1mm}

\Paragraph{\Ob Full Leveling has the Highest Mean Compaction Latency} 
As expected, \texttt{Full} compactions have the highest average latency 
($1.2$--$1.9\times$ higher than partial leveling, and $2.1\times$ than tiering).
The mean compaction latency is observed to be directly proportional to the average amount of data moved per compaction. 
\texttt{Full} can neither take advantage of pseudo-compactions nor optimize the data movement during compactions, hence, on average the data movement per compaction remains large. 
\texttt{1-Lvl} provides the most predictable performance in terms of compaction latency.
Fig. \ref{fig:W1}(c) shows the mean compaction latency for all
strategies as well as the median (P50), the $90^{th}$ percentile (P90), 
the $99^{th}$ percentile (P99), and the maximum (P100). 
The tail compaction latency largely depends on the amount of data moved by the largest compaction jobs triggered during the workload execution. 
We observe that the tail latency (P90, P99, P100) is more predictable for \texttt{Full}, while partial compactions, and especially, tiering have high variability due to differences in the data movement policies.
% and the key-overlap depending on the data distribution.

The compaction latency presented in Fig. \ref{fig:W1}(c) can be broken to
IO time and CPU time. We observe that the CPU effort is about $50\%$ 
regardless of the compaction strategy.
During a compaction, CPU cycles are spent in (1) obtaining locks and taking  
snapshots, (2) merging the entries, (3) updating file 
pointers and metadata, and (4) synchronizing output files post compaction. 
Among these, the time spent to sort-merge the data in memory dominates.

\begin{figure*}[!ht]
\vspace{-0.3in}
\centering
\includegraphics[width=1.02\textwidth]{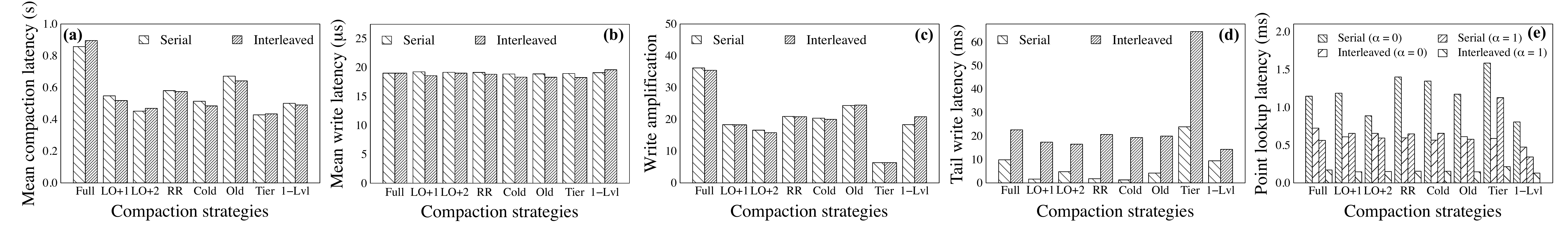}
\vspace*{-0.35in}
\caption{(a-c) The average ingestion performance for workloads with interleaved inserts and queries is similar to that of an insert-only workload, but (d) with worse tail performance. However, (e) interleaved lookups are significantly faster.} 
\vspace*{-0.1in}
% (a) rocksdb.compact.read.bytes (231) + rocksdb.compact.write.bytes (232); (b) rocksdb.compaction.times.micros (321); (c) rocksdb.compaction.times.cpu\_micros (327); (d) rocksdb.db.write.micros (315); (e) rocksdb.db.get.micros (309); (f) rocksdb.db.seek.micros (429)
\label{fig:W1-mixed}
\end{figure*}

\Paragraph{The Tail Write Latency is Highest for Tiering} 
Fig. \ref{fig:W1}(d) shows that the tail write latency is highest for tiering. 
The tail write latency for \texttt{Tier} is $\sim$$2.5\times$ greater than \texttt{Full} and $5$--$12\times$ greater than partial compactions. 
Tiering in RocksDB~\cite{RocksDB2020} optimizes for writes and opportunistically seeks to compact all data to a large single level. 
This design achieves lower average write latency (Fig. \ref{fig:W1-mixed}(b)) at the expense of prolonged write stalls in the worst case, which is when the overlap between two consecutive levels is very high. 
\texttt{Full} also has $2$--$5\times$ higher tail write stalls than partial compactions because when multiple consecutive levels are close to saturation, a buffer flush can result in a cascade of compactions. 
\texttt{1-Lvl} too has a higher tail write latency as the first level is realized as tiering.

\vspace{1mm}
\noindent\fbox{%
    \parbox{0.465\textwidth}{%
    \small
        \Paragraph{\TA \textit{\texttt{Tier} may cause prolonged write stalls}} 
        \textit{Tail write stall for \texttt{Tier} is $\sim$$25$ms, while for partial leveling (\texttt{Old}) it is as low as $1.3$ms.} 
    }%
    \normalsize
}
\vspace{1mm}

% \vspace{-0.1in}
\subsubsection{\textbf{Querying the Data}}
In this experiment, we perform $1$M point lookups on the previously generated 
preloaded database (with $10$M entries). The lookups are uniformly distributed
in the domain and we vary the fraction of empty lookups $\alpha$ between 0 and 1.
Specifically, $\alpha = 0$ indicates that we consider only non-empty lookups, 
while for $\alpha = 1$ we have lookups on non-existing keys. We also execute
$1000$ range queries, while varying their selectivity.

\Paragraph{\Ob The Point Lookup Latency is Highest for Tiering and Lowest for Full-Level Compaction}
Fig. \ref{fig:W1}(e) shows that point lookups perform the best for \texttt{Full}, and the worst for tiering. 
The mean latency for point lookups with tiering is between 
$1.1$--$1.9\times$ higher than that with leveled compactions for lookups on existing keys, and $\sim$$2.2\times$ higher for lookups on non-existing keys. 
Note that lookups on existing keys must always perform at least one I/O per 
lookup (unless they are cached).
%, and therefore, takes longer than that on non-existing keys. 
%
For non-empty lookups in a tree with size ratio $T$, theoretically,
the lookup cost for tiering should be $T\times$ higher than its leveling 
equivalent~\cite{Dayan2017}. 
However, this \textit{worst-case} cost is not always accurate; in practice it depends on (i) the block cache size and the caching policy, (ii) the temporality of the lookup keys, and (iii) the implementation of the compaction strategies. 
RocksDB-tiering has overall fewer sorted runs than textbook tiering.
Taking into account the block cache and temporality in the lookup workload,
the observed tiering cost is less than $T\times$ the cost observed for 
\texttt{Full}. In addition, \texttt{Full} is $3\%$--$15\%$ lower than the partial 
compaction routines, because during normal operation of \texttt{Full} some levels
might be entirely empty, while for partial compaction all levels are always 
close to being full.
Finally, we note that the choice of data movement policy does not affect the
point lookup latency significantly, which always benefits from Bloom filters 
($10$ bits-per-key) and the block cache ($0.05\%$ of the data size).

\Paragraph{Point Lookup Latency Increases for Comparable Number of Empty and Non-Empty Queries} A surprising result for point lookups 
that is also revealed in Fig. \ref{fig:W1}(e) is that they perform worse when the 
fraction of empty and non-empty lookups is balanced.
Intuitively, one would expect that as we have more empty queries (that is, as $
\alpha$ increases) the latency would decrease since the only data accesses needed 
by empty queries are the ones due to Bloom filter false 
positives~\cite{Dayan2017}.
To further investigate this result, we plot in 
Fig.~\ref{fig:W1}(h) the $90^{th}$
percentile ($P90$) latency which shows a similar
curve for point lookup latency as we vary $\alpha$. 
%We further plot the block cache misses for Bloom filters 
%blocks in Fig.~\ref{fig:W1}(h2), and the index (fence 
%pointer) block misses in Fig.~\ref{fig:W1}(h3). 
%
In our configuration each file uses $20$ pages
for its Bloom filters, $4$ pages for its index blocks, and 
that the false positive is $FPR=0.8\%$. 
A non-empty query needs to load the Bloom filters of the levels it visits until it terminates. 
For all intermediate levels, it accesses the index and data blocks with probability $FPR$, and then fetches the index and data blocks for the target level. 
On the other hand, an empty query probes the Bloom filters of all levels before returning
an empty result. Note that for each level
it also accesses the index and data blocks with $FPR$.
The counter-intuitive shape is a result of the
non-empty lookups not needing to load the Bloom filters
for all levels when $\alpha=0$ and the 
empty lookups
accessing index and data only when there is a false 
positive when $\alpha=1$.
Fig. \ref{fig:W1}(h) also shows the highly predictable point lookup performance of \texttt{1-Lvl}. 

%\red{ 
%Both empty and non-empty lookups must first fetch the filter blocks, hence,
%for the filter block misses remain almost unaffected as we vary $\alpha$.
%Not that \texttt{Tier} has more misses because it has more overall sorted runs. 
%Subsequently, the index blocks are fetched only if the filter probe returns 
%positive. With $10$ bits-per-key the false positive is only 0.8\%, and as we have 
%more empty queries, that is, increasing $\alpha$, fewer index blocks are 
%accessed. The filter blocks are maintained at a granularity of files and in our
%setup amount to $20$ I/Os. The index blocks are maintained for each disk page
%and in our setup amount to $4$ I/Os, being $1/5^{th}$ of the cost for fetching
%the filter blocks. 
%%\footnote{filter block size per file = \#entries per file $*$ bits-per-key = $512$*$128$*$10$B = $80$kB; index block size per file = \#entries per file $*$ (key size$+$pointer size) = $512 * (16$+$16)$B = $16$kB.}. 
%%The cost for fetching the filter block is $5\times$ the cost for fetching the index block. 
%This, coupled with the probabilistic fetching of the index block (depending on $\alpha$ and the false positive rate ($FPR=0.8\%$) of the filter) leads to a non-monotonic latency curve for point lookups as $\alpha$ increases, and this behavior is persistent regardless of the compaction strategy.} 

\vspace{2mm}
\noindent\fbox{%
    \parbox{0.465\textwidth}{%
    \small
        \Paragraph{\TA \textit{The point lookup latency is largely unaffected by the data movement policy}} \textit{In presence of Bloom filters (with high enough memory) and small enough block cache, the point query latency remains largely unaffected by the data movement policy as long as the number of sorted runs in the tree remains the same. This is because block-wise caching of the filter and index blocks reduces the time spent performing disk I/Os significantly.} 
    }%
    \normalsize
}
\vspace{0.7mm}

%\vspace{-0.04in}
\Paragraph{\Ob Read Amplification is Influenced by the Block Cache Size and File Structure, and is Highest for Tiering} 
Fig. \ref{fig:W1}(f) shows that the read amplification across different 
compaction strategies for non-empty queries ($\alpha=0$) is between $3.5$ and 
$4.4$. This is attributed to the size of filter and index blocks which 
are $5\times$ and $1\times$ the size of a data block, respectively. 
Each non-empty point lookup fetches between $1$ and $L$ filter blocks depending 
on the position of the target key in the tree, and up to $L \cdot FPR$
%\footnote{$FPR$ is estimated as $e^{- \tfrac{M_{BF}}{N} \cdot (\log{2})^2}$, where $M_{BF}$ is the memory assigned for Bloom filters and $N$ is the number of entries in the tree.} 
index and data blocks. 
Further, the read amplification increases exponentially with $\alpha$, reaching up to $14.4$ for leveling and $21.3$ for tiering (for $\alpha=0.8$).
%This is because each empty point lookup must fetch $L$ filter blocks (in addition to any index and data block), and any disk I/Os performed during an empty point lookups is deemed as superfluous. 
Fig. \ref{fig:W1}(f) also shows that the estimated read amplification for point lookups is between $1.2\times$ and $1.8\times$ higher for \texttt{Tier} than for leveling strategies. 
This higher read amplification for \texttt{Tier} is owing to the larger number of sorted runs in the tree, and is in line with \textbf{O4}.

\Paragraph{The Effect of Compactions on Range Scans is Marginal} 
To answer a range query, LSM-trees instantiate multiple \textit{run-iterators} scanning all sorted runs containing qualifying data.
Thus, its performance depends on (i) the iterator scan time (which relates to selectivity) and (ii) the time to merge the data. 
The number of sorted runs in a 
leveled LSM-tree remains the same, which results in similar range query latency for all leveled variations, especially for larger selectivity (Fig. \ref{fig:W1}(g)). 
Note that without updates or deletes, the amount of data qualifying for a range query remains largely identical for different data layouts despite the number of runs being different. 
The $\sim$$5\%$ higher average range query latency for \texttt{Tier}
is attributed to the additional I/Os needed to handle partially
qualifying disk pages from each run ($O(L\cdot T)$ in the worst case).

% \vspace{2mm}
% \noindent\fbox{%
%     \parbox{0.465\textwidth}{%
%     \small
%         \Paragraph{\TA \textit{In absence of updates/deletes, the range query latency is unaffected by compactions}} \textit{For workloads with unique inserts, the total amount of data sort-merged by the iterators is largely similar for all compaction strategies, leading to similar range query performance.} 
%     }%
%     \normalsize
% }
%\vspace{1mm}

% \vspace{-0.1in}
\subsubsection{\textbf{Executing mixed workloads}}
We now discuss the performance implications when ingestion and queries are mixed. 
We interleave the ingestion of $10$M 
unique key-value entries with $1$M point lookups. 
The ratio of empty to non-empty lookups is varied across experiments.
All lookups are performed after $L-1$ levels are full. 
Fig.~\ref{fig:W1-mixed} compares side by side the results for serial and interleaved execution of workloads with same specifications.
%Unless otherwise mentioned, the point lookups are 
%performed on existing keys.

\begin{figure*}[t]
	\vspace{-0.3in}
    \centering
%    \vspace*{-0.2in}
    \includegraphics[width=0.95\textwidth]{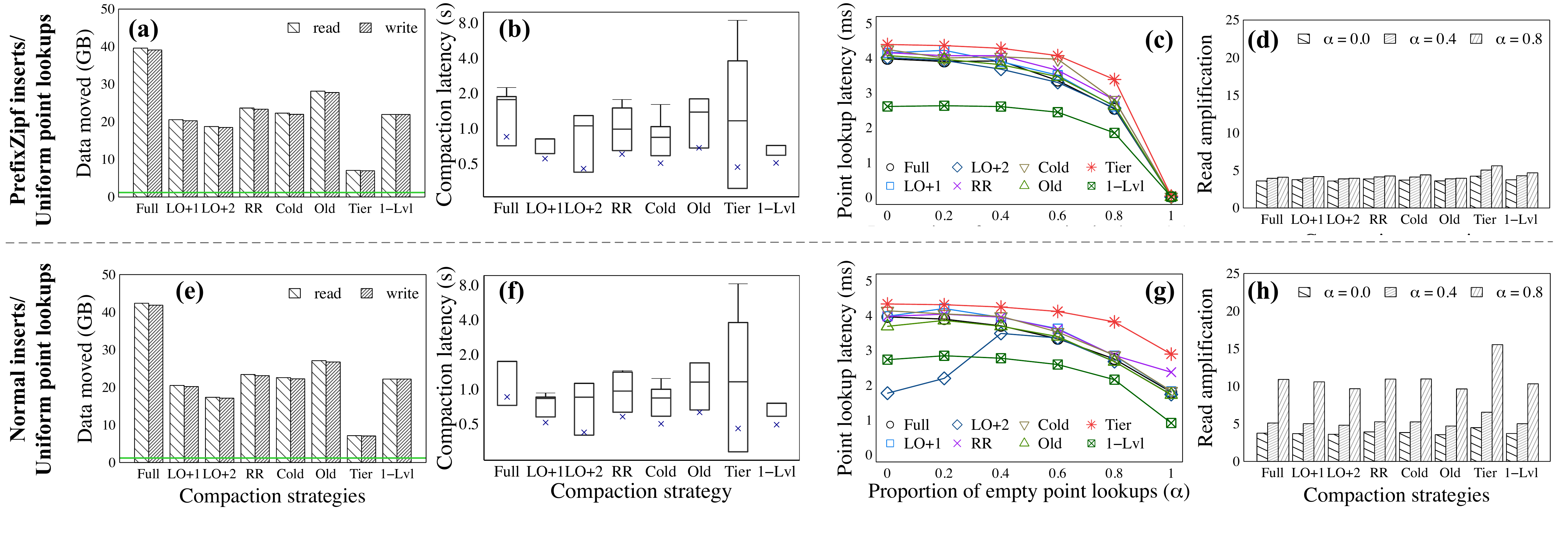}
    \vspace*{-0.25in}
    \caption{As the ingestion distribution changes to (a-d) PrefixZipf and (e-h) normal with standard deviation, the ingestion performance of the database remains nearly identical with improvement in the lookup performance.} 
    \vspace*{-0.1in}
    % (a) rocksdb.compact.read.bytes (231) + rocksdb.compact.write.bytes (232); (b) rocksdb.compaction.times.micros (321); (c) rocksdb.compaction.times.cpu\_micros (327); (d) rocksdb.db.write.micros (315); (e) rocksdb.db.get.micros (309); (f) rocksdb.db.seek.micros (429)
    \label{fig:W5}
    \end{figure*}
    
    \begin{figure*}[h]
        \centering
        \includegraphics[width=0.95\textwidth]{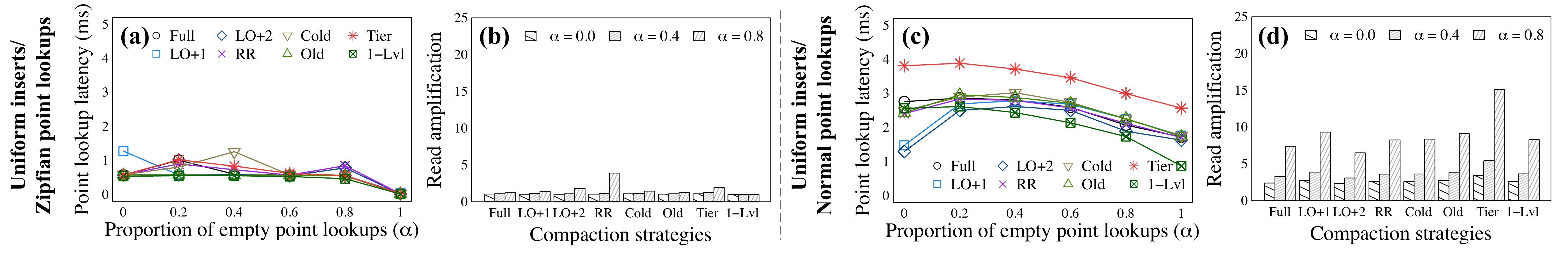}	
        % \caption{\#hit of filter blocks}
        \vspace*{-0.25in}
        \caption{Skewed lookup distributions like Zipfian (a, b) and normal (c, d) improve the lookup performance dramatically in the presence of a block cache and with the assistance of Bloom filters.} 
        \vspace*{-0.1in}
        \label{fig:W6}
    \end{figure*}

\Paragraph{\Ob Mixed Workloads have Higher Tail Write Latency} 
Figures~\ref{fig:W1-mixed}(a) and (b) show that the
mean latency of compactions that are interleaved with
point queries is only marginally affected for all 
compaction strategies. This is also corroborated by the
write amplification remaining unaffected by mixing
reads and writes as shown in Fig.~\ref{fig:W1-mixed}(c).
On the other hand, Fig.~\ref{fig:W1-mixed}(d) shows that 
the tail write 
latency is increased between $2$--$15\times$.
This increase is attributed to (1) the need of point
queries to access filter and index blocks that requires
disk I/Os that compete with writes and saturate the 
device, and (2) the delay of
memory buffer flushing during lookups.

\Paragraph{Interleaving Compactions and Point Queries Helps Keeping the Cache Warm} 
Since in this experiment we start the point queries when
$L-1$ levels of the tree are full, we expect that the 
interleaved read query execution will be faster than
the serial one, by $1/L$ (25\% in our configuration) which 
corresponds to the difference in the height of the trees.
However, Fig. \ref{fig:W1-mixed}(e) shows this 
difference to be between $26\%$ and $63\%$ for non-empty 
queries and between $69\%$ and $81\%$ for empty queries.  
The reasons interleaved point query execution is faster
than expected are that (1) about $10\%$ of lookups 
terminate within the memory buffer, without requiring any 
disk I/Os, and (2) the block cache is pre-warmed with 
filter, index, and data blocks cached during compactions. 
%
%Putting together the marginal impact of interleaved 
%execution to compactions and the significant benefits for
%point queries, the total workload execution time is
%$1.6$--$2.1\times$ faster than serial execution as shown in Fig. \ref{fig:W1-mixed}(f).
Fig. \ref{fig:W1-mixed}(d) and \ref{fig:W1-mixed}(e) show how \texttt{1-Lvl} brings together \textit{the best of both worlds} and offer reasonably good ingestion and lookup performance simultaneously.

%This is due to the superior lookup performance in mixed workloads, as the mean compaction latency, the write latency, as well as the write amplification are comparable for serial and interleaved workloads, as shown in Figures \ref{fig:W1-mixed}(b), \ref{fig:W1-mixed}(c), and \ref{fig:W1-mixed}(e), respectively. 
%While the lookups in a mixed workload are executed on a samller-sized database with one fewer levels at the beginning, it is expected that the mean point lookup latency would be $\sim$$3\%$ and $\sim$$12.5\%$ less than that of a lookup-only workload for non-empty and empty point queries, respectively. 

%However, Figure \ref{fig:W1-mixed}(f) shows this difference to be between $26\%$ and $63\%$ for non-empty queries and between $69\%$ and $81\%$ for empty queries.  
%Point lookups are realized significantly faster for a mixed workload as: (1) about $10\%$ of lookups terminate within the memory buffer, without requiring any disk I/Os; (2) temporality of the lookup workload also increases the hits to Level $1$ and Level $2$, which are often cached in memory; and (3) the block cache is pre-warmed with filter, index, and data blocks cached during compactions. 
%However, when point lookups are performed on non-existing keys, the mean lookup latencies are comparable for mixed and query-only workloads as (1) there are no early termination for a lookup in either case, and (2) the number of expected disk I/Os to fetch the data blocks is very low in both cases. 

\vspace{1mm}
\noindent\fbox{%
    \parbox{0.465\textwidth}{%
    \small
        \Paragraph{\TA \textit{Compactions help lookups by warming up the caches}} \textit{As the file metadata is updated during compactions, the block cache is warmed up with the filter, index, and data blocks, which helps subsequent point lookups.} 
    }%
    \normalsize
}
\subsection{Workload Influence}
\label{subsec:workload}
Next, we analyze the implications of the workloads on compactions.

%\Paragraph{Setup} Default.  

% \vspace{-0.1in}
\subsubsection{\textbf{Varying the Ingestion Distribution}}
In this experiment, we use an interleaved workload that varies the ingestion distribution (\textit{Zipfian} with $s=1.0$, \textit{normal} with $34\%$ standard deviation), and has uniform lookup distribution. 
We use a variant of the Zipfian distribution, called \emph{PrefixZipf}, where the key prefixes follow a Zipfian distribution while the suffixes are generated uniformly. 
This allows us to avoid having too many updates in the workload.

\Paragraph{Ingestion Performance is Agnostic to Insert Distribution} 
Figures \ref{fig:W1}(a), \ref{fig:W5}(a), and \ref{fig:W5}(e) show that the total data movement during compactions remains virtually identical for (unique) insert-only workloads generated using uniform, PrefixZipf, and normal distributions, respectively. 
Further, we observe that the mean and tail compaction latencies are agnostic of the ingestion distribution
(Fig. \ref{fig:W1}(c), \ref{fig:W5}(b), and \ref{fig:W5}(f) are almost identical as well). 
As long as the data distribution does not change over
time, the entries in each level follow the
same distribution and the overlap between 
different levels remains the same.
\textit{Therefore, for an ingestion-only workload 
the data distribution does not influence the 
choice of compaction strategy.} 

\Paragraph{\Ob Insert Distribution Influences Point Queries} 
Figure \ref{fig:W5}(c) shows that while tiering has a 
slightly higher latency for point lookups, the relative 
performance of the compaction strategies is close to each 
other for any fraction of non-empty queries in the 
workload (all values of $\alpha$). 
This is because when empty queries are drawn uniformly from the key domain, the level-wise metadata and index blocks help to entirely avoid a vast majority of unnecessary disk accesses (including fetching index or filter blocks). \
%In presence of Prefix Bloom filters~\cite{RocksDB2020b}, as $\alpha$$\to$$1$, the filter and index block misses in the block cache diminishes sharply and approaches zero. 
%Note that in this experiment the inserts are generated using PrefixZipf distribution and target a small part of the domain. 
%This allows a large fraction of the non-existing 
%point lookups to terminate in memory by simply utilizing 
%file metadata (i.e., the min/max keys contained in each 
%file) without requiring to fetch any filter (and index) 
%blocks.
In Fig.~\ref{fig:W5}(d), we observe that the read 
amplification remains comparable to that in 
Fig. \ref{fig:W1}(f) (\textit{uniform} ingestion) for $\alpha = 0$ and even $\alpha = 0.4$. 
However, for $\alpha = 0.8$, the read amplification in 
Fig. \ref{fig:W5}(d) becomes $65\%$-$75\%$ smaller than
in the case of uniform inserts. The I/Os performed to 
fetch the filter blocks is close to zero. 
\textit{This shows that all compaction strategies perform equally well while executing an empty query-heavy workload on a database pre-populated with PrefixZipf inserts.}
In contrast, when performing lookups on a database pre-loaded with normal ingestion, the point lookup performance (Fig. \ref{fig:W5}(g)) largely resembles its uniform equivalent (Fig. \ref{fig:W1}(h)), as the ingestion-skewness is comparable. 
The filter and index block hits are $\sim10\%$ higher for the normal distribution compared to uniform for larger values of $\alpha$, which explains the comparatively lower read amplification shown in Fig. \ref{fig:W5}(h). 
This plot also shows the first case of unpredictable behavior of \texttt{LO+2} for $\alpha=0$ and $\alpha=0.2$. 
We observe more instances of such unpredictable behavior for \texttt{LO+2}, which probably explains why it is rarely used in new LSM stores. 
Once again, for both the compaction and tail lookup performance, \texttt{1-Lvl} offers highly predictable performance.

% \Paragraph{\mob Range Queries are Marginally Affected by the Insert Distribution} 
% \red{re-plotting needed!}
 
% \vspace{-0.1in}
\subsubsection{\textbf{Varying the Point Lookup Distribution}}
In this experiment, we change the point lookup 
distribution to \textbf{Zipfian} and \textbf{normal}, 
while keeping the ingestion distribution as uniform. 

\Paragraph{The Distribution of Point Lookups Significantly Affects Performance}
Zipfian point lookups on uniformly populated data
leads to low latency point queries for all compaction 
strategies, as shown in Fig. \ref{fig:W6}(a) because the 
block cache is enough for the popular blocks in all cases, 
as also shown by the low read amplification in 
Fig. \ref{fig:W6}(b).
On the other hand, when queries follow the normal
distribution, partial compaction strategies \texttt{LO+1} 
and \texttt{LO+2} dominate all other approaches, while 
\texttt{Tier} is found to perform significantly 
slower than all other approaches, as shown in 
Fig. \ref{fig:W6}(c) and \ref{fig:W6}(d). 

\vspace{1mm}
\noindent\fbox{%
    \parbox{0.465\textwidth}{%
    \small
        \Paragraph{\TA \textit{For skewed ingestion/lookups, all compaction strategies behave similarly in terms of lookup performance}} \textit{While the ingestion distribution does not influence its performance, heavily skewed ingestion or lookups impacts query performance due to block cache and file metadata.} 
    }%
    \normalsize
}
\vspace{1mm}

\begin{figure*}[h]
	\vspace{-0.3in}
    \centering
    \includegraphics[width=\textwidth]{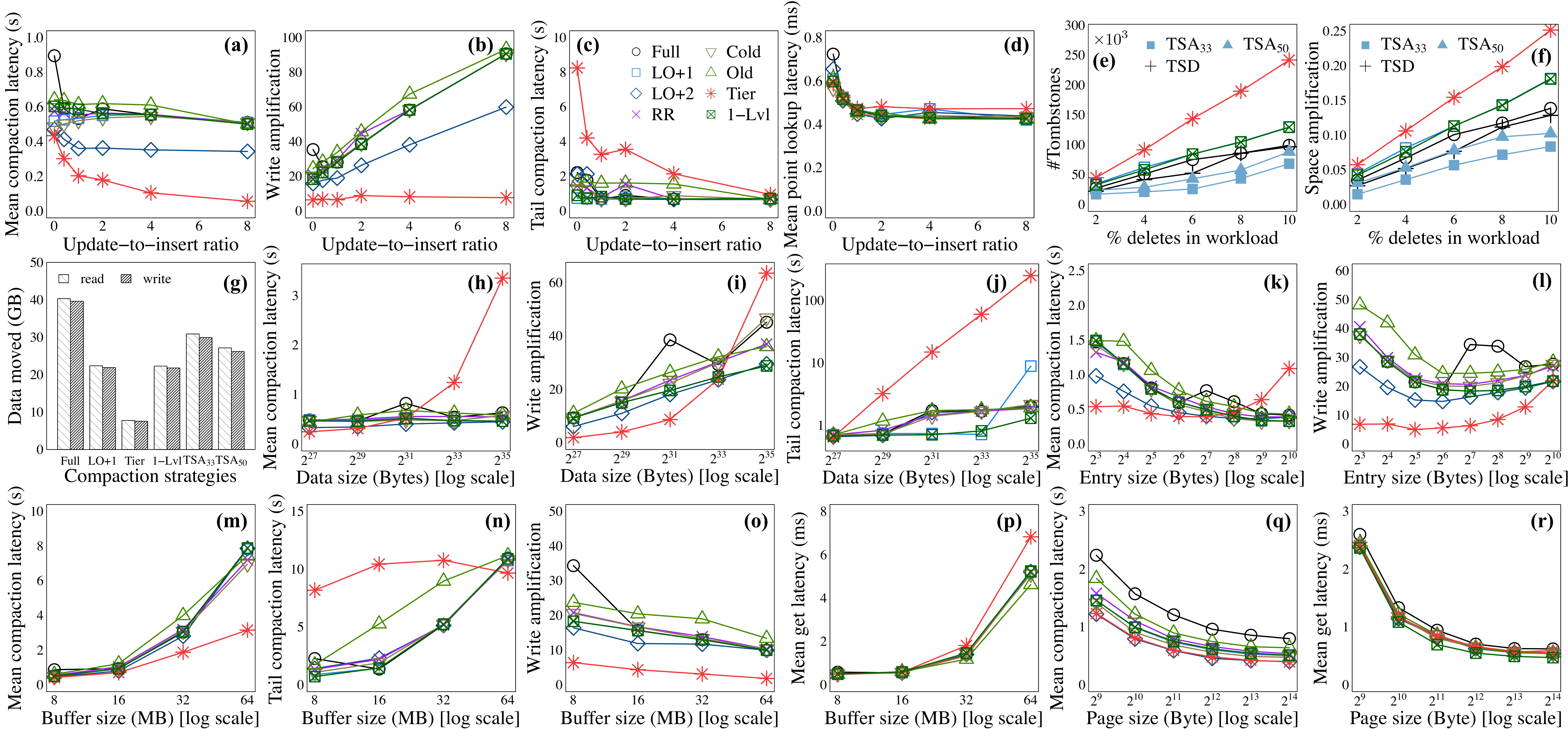}
    \vspace*{-0.2in}
    \caption{Experiments with varying workload and data characteristics (a-l) and LSM tuning (m-r) show that there is no perfect compaction strategy -- choosing the appropriate compaction strategy is subject to the workload and the performance goal.} 
    \label{fig:W7}
    \vspace*{-0.1in}
\end{figure*}

% \vspace{-0.1in}
\subsubsection{\textbf{Varying the Proportion of Updates}}
We now vary the update-to-insert ratio, while interleaving
queries with ingestion. An update-to-insert ratio $0$ means 
that all inserts are unique, while a ratio $8$ means that 
each unique insert receives $8$ updates on average.

\Paragraph{\Ob For Higher Update Ratio Compaction Latency for Tiering Drops; \texttt{LO+2} Dominates the Leveling Strategies}
As the fraction of updates increases, the mean compaction latency decreases
significantly for tiering because we discard multiple
updated entries in every compaction (Fig. \ref{fig:W7}(a)). 
We observe similar
but less pronounced trends for \texttt{Full} and 
\texttt{LO+2}, while the remaining leveling strategies
remain largely unchanged. 
\textit{Overall, larger 
compaction granularity helps to exploit the presence of 
updates by invalidating more entries at a time.} 
Among the leveling strategies, \texttt{LO+2} performs best as
it moves $\sim$$20\%$ less 
data during compactions, which also affects write 
amplification as shown in Fig. \ref{fig:W7}(b).

%Also, previously we discussed that \texttt{Tier} has the highest tail compaction latency. 
As the fraction of updates increases, all compaction strategies including
\texttt{Tier} have lower tail compaction latency.  
Fig. \ref{fig:W7}(c) shows that \texttt{Tier}'s tail compaction latency
drops from $6\times$ higher than \texttt{Full} to $1.2\times$ for an 
update-to-insert ratio of $8$, which demonstrates that \texttt{Tier} is
most suitable for update-heavy workloads. We also observe that lookup
latency and read amplification also decrease for update-heavy workloads.

\Paragraph{The Point Lookup Latency Stabilizes with the Level Count}
Fig. \ref{fig:W7}(d) shows that as the update-to-insert ratio increases, the mean point lookup latency decreases sharply before stabilizing. 
The initial sharp fall in the latency is attributed to a decrement in the number of levels (from $4$ to $3$) in the LSM-tree, when the update-to-insert ratio increases from $0.4$ to $1$. 
The latency then stabilizes because non-empty point lookups perform at least one disk I/O, which, in turn, dominates the overall lookup cost.

%\Paragraphit{\mob Point Lookup Latency, Read Amplification, and Range Lookup Latency all Decrease with Increase in Updates Similarly for all Compaction Strategies} 

\vspace{1mm}
\noindent\fbox{%
    \parbox{0.465\textwidth}{%
    \small
        \Paragraph{\TA \textit{Tiering dominates the performance for update-intensive workloads}} \textit{When subject to update-intensive workloads, \texttt{Tier} exhibits superior compaction performance  along with comparable lookup performance (as leveled LSMs), which allows it to dominate the overall performance space.} 
    }%
    \normalsize
}
\vspace{1mm}

% \vspace{-1mm}
\subsubsection{\textbf{Varying Delete Proportion}}
We now analyze the impact of deletes, which manifest as out-of-place 
invalidations with special entries called tombstones \cite{Sarkar2020}.
We keep the same data size and vary the proportion of point 
deletes in the workload. All deletes are issued on existing keys and are 
interleaved with the inserts.

\Paragraph{\texttt{TSD} and \texttt{TSA} Offer Superior Delete Performance} 
We quantify the efficacy of deletion using the number of tombstones at the end
of the workload execution. 
The lower this number, the faster deleted data has
been purged from the database, which in turn reduces space, write, and read
amplification. 
Fig. \ref{fig:W7}(e) shows that \texttt{TSD} and \texttt{TSA} maintain the fewer tombstones at the end of the experiment.
For a workload with $10\%$ deletes, \texttt{TSD} purges $16\%$ more tombstones than \texttt{Tier} and $5\%$ more tombstones than \texttt{LO+1} by picking the files that have a tombstone density above a pre-set threshold for compaction. 
For \texttt{TSA}, we experiment with two different thresholds for delete persistence: \texttt{TSA}$_{33}$ and \texttt{TSA}$_{50}$  is set to $33\%$ and $50\%$ of the experiment run-time, respectively. 
As \texttt{TSA} guarantees persistent deletes within the thresholds set, it compacts more data aggressively, and ends up with $7$--$10\%$ fewer tombstones as compared to \texttt{TSD}. 
\texttt{Full} manages to purge more tombstones than any partial compaction routine, as it periodically compacts entire levels.
\texttt{Tier} retains the highest number of tombstones as it maintains the highest number of sorted runs overall. 
As the proportion of deletes in the workload increases, the number of tombstones remaining the LSM-tree (after the experiment is over) increases. 
\texttt{TSA} and \texttt{TSD} along with \texttt{Full} scale better than the partial compaction routines and tiering. 
By compacting more tombstones, \texttt{TSA} and \texttt{TSD} also purge more invalid data reducing space amplification, as shown in Fig. \ref{fig:W7}(f).

\Paragraph{\Ob Optimizing for Deletes Comes at a (Write) Cost}
The reduced space amplification offered by \texttt{TSA} and \texttt{TSD} is achieved by compacting the tombstones eagerly, which increases the overall amount of data moved due to compaction. 
Fig. \ref{fig:W7}(g) shows that \texttt{TSD} and \texttt{TSA}$_{50}$ compacts 
$18\%$ more data than the write optimized \texttt{LO+1} (for \texttt{TSA}$_{33}$ this becomes $35\%$). 
Thus, \texttt{TSD} and \texttt{TSA} are useful when the objective is to (i) persist deletes timely or (ii) reduce space amplification caused by deletes.

\vspace{1mm}
\noindent\fbox{%
    \parbox{0.465\textwidth}{%
    \small
        \Paragraph{\TA \textit{\texttt{TSD} and \texttt{TSA} are tailored for deletes}} \textit{\texttt{TSA} and \texttt{TSD}, by design, choose files with tombstones for compactions to reduce space amplification. \texttt{TSA} ensures timely persistent deletion by compacting more data eagerly for smaller persistence thresholds, which increases the write amplification.} 
    }%
    \normalsize
}
\vspace{1mm}

% \vspace{-1mm}
\subsubsection{\textbf{Varying the Ingestion Count}}
We now report the scalability results by varying the data size from $2^{27}$B to $2^{35}$B.

\Paragraph{\Ob \texttt{Tier} Scales Poorly Compared to Leveled and Hybrid Strategies} 
The mean compaction latency scales sub-linearly for all compaction strategies barring \texttt{Tier}, as shown in Fig. \ref{fig:W7}(h).
\textit{The relative advantages of compaction strategies with leveled and hybrid data layouts remain similar regardless of the data size.}
This observation is further backed up by Fig. \ref{fig:W7}(i) which shows how write amplification scales. 
We also observe that the advantages of the RocksDB-implementation of tiering (i.e., \textit{universal compaction})~\cite{RocksDB2020} diminishes as the data size grows beyond $8$GB. 
Fig.~\ref{fig:W7}(j) shows that as the data size increases, the tail compaction latency for \texttt{Tier} increases, as the worst-case overlap between files from consecutive levels increase significantly.
This makes \texttt{Tier} unsuitable for latency-sensitive applications. 
%Fig.~\ref{fig:W7}(j) shows that as the data size grows, the worst-case overlap between files from consecutive levels increases significantly for \texttt{Tier} which increases the tail compaction latency, rendering \texttt{Tier} unsuitable for latency-sensitive applications. 
When the data size reaches $2$GB, \texttt{Full} triggers a \textit{cascading compaction} that writes all data to a new level, causing spikes in write amplification and compaction latency.

% \vspace*{-1mm}
\subsubsection{\textbf{Varying Entry Size}}
Here, we keep the key size constant ($4$B) and vary the value from $4$B to 
$1020$B to vary the entry size. 

\Paragraph{\Ob For Smaller Entry Size, Leveling Compactions are More Expensive} 
%The number of key-value entries to be merged is a key factor of the cost of 
%a compaction. 
Smaller entry size increases the number of 
entries per page, which in turn, leads to (i) more keys to be compared during
merge and (ii) bigger Bloom filters that require more space per file and more CPU for hashing. 
Fig. \ref{fig:W7}(k) shows these trends. We also observe similar
trends for write amplification in Fig. \ref{fig:W7}(l) and for query latency. 
They both decrease as the entry size increases. 
However, as the overall data size increases with the entry size, we observe the compaction latency and write amplification to increase steeply for \texttt{Tier} (similarly to Fig. \ref{fig:W7}(h) and (i)).

%that for smaller key-value size, the mean compaction latency leveling compaction strategies exhibit a higher mean compaction latency. 
%This is because the memory buffer holds more entries when the entry size is small, which is also the case for the files on disk. 
%Therefore, during each compaction, the number of entries compacted is higher when the entry size is smaller. 
%This adds to the compaction latency as: (1) the number of user-key comparisons per compaction increases, (2) the number of Bloom filter hashes per file becomes higher and (3) a larger size of the filter block requires more disk I/Os to be done. 
%
%\Paragraph{\mob As the Entry Size Increases, the Compaction Latency as well as the Point Lookup Latency Decreases} 
%
%\Paragraph{\mob For a Constant Page Size, Read Amplification Reduces as the Entry Size Increases} 

% ############## Section 5.3 ##############
%\vspace{-0.1in}
\subsection{LSM Tuning Influence}
\label{subsec:tuning}
In the final part of our analysis, we discuss the interplay of  
compactions with the standard LSM tunings knobs, such as memory buffer size, page size, and size ratio.
%, and analyze their impact on the compaction strategies employed. 
%In the interest of space, we only include the most important results, while the rest can be found in an extended version of this paper~\cite{Sarkar2021b}.
%
%\subsubsection{Varying the Memory Buffer Size}

\Paragraph{\Ob Compactions with Tiering Scale Better with Buffer Size} 
Fig. \ref{fig:W7}(m) shows that as the buffer size increases, the mean compaction latency increases across all compaction strategies. 
The size of buffer dictates the size of the files on disk, and 
larger file size leads to more data being moved per compaction.
Also, for larger file size, the filter size per file increases along with the time spent for hashing, which increases compaction latency. 
Further, as the buffer size increases, the mean compaction latency for \texttt{Tier} scales better than the other strategies.
Fig. \ref{fig:W7}(n) shows that the high tail compaction latency for \texttt{Tier} plateaus quickly as the buffer size increases, and eventually crossovers with that for the eagerer compaction strategies when the buffer size becomes $64$MB. 

We also observe in Fig. \ref{fig:W7}(o) that among the partial compaction routines \texttt{Old} experiences an increased write amplification throughout, while \texttt{LO+1} and \texttt{LO+2} consistently offer lower write amplification and guarantee predictable ingestion performance. 
Fig. \ref{fig:W7}(p) shows that as the memory buffer size increases, the mean point lookup latency increases superlinearly. 
This is because, for larger memory buffers, the files on disk hold a greater number of pages, and thereby, more entries. 
Thus, the combined size of the index block (one index per page) and filter block (typically, $10$ bits per entry) per file grows proportionally with the memory buffer size. 
The time elapsed in fetching the index and filter blocks causes the mean latency for point lookups to increase significantly. 

\Paragraph{All Compaction Strategies React Similarly to Varying the Page Size} 
In this experiment, we vary the logical page size, which in turn, changes the number of entries per page. 
The smaller the page size, the larger the number of pages per file -- meaning more I/Os are required to access a file on the disk. 
For example, when the page size shrinks from $2^{10}$B to $2^9$B, the number of pages per file doubles. 
With smaller page size, the index block size per file increases as more pages should be indexed, which also contributes to the increasing I/Os. 
Thus, an increase in the logical page size, reduces the mean compaction latency, as shown in Fig.~\ref{fig:W7}(q). 
In Fig.~\ref{fig:W7}(r), we observe that as the page size increases, the size of the index block per file decreases, and on average fewer I/Os are performed to fetch the metadata block overall for every point lookup. 

\Paragraph{Miscellaneous Observations} 
We also vary LSM tuning parameters such as the size ratio, the memory allocated 
to Bloom filters, and the size of the block cache. 
We observe that changing the values of these knobs
affects the different compaction strategies similarly, and 
hence, does not influence the choice of the appropriate 
compaction strategy for any particular set up. 
\section{Discussion}
%\vspace{-0.02in}
\label{sec:tuning}
The design space detailed in Section \ref{sec:compaction} and 
the experimental analysis presented in Section \ref{sec:results}
aim to offer to database researchers and practitioners the necessary insights
to make educated decisions when selecting compaction strategies for LSM-based data stores.
%We use the term \textit{appropriate}, instead of \textit{optimal} or \textit{best}, as our goal is to take informed decisions to \textbf{avoid the worst choices} as opposed to \textit{finding the optimal/best choice}. 

% will also enhance our conventional wisdom about the impact of compaction on LSM-based storage engines.

\Paragraph{Know Your LSM Compaction}
LSM-trees are considered ``write-optimized'', however, in practice their performance strongly depends on \textit{when and how compactions are performed}. 
We depart from the notion of \textit{treating compactions as a 
black-box}, and instead, we formalize \textit{LSM compactions} 
as an ensemble of four fundamental compaction primitives. 
This allows us to reason about each of these primitives and 
navigate the \textit{LSM compaction design space} in search of 
the appropriate compaction strategy for a workload or for 
custom performance goals. 
Further, the proposed compaction design space provides the
necessary intuitions about how simple modifications (like data movement policy or compaction granularity) to an existing engine (like RocksDB) can be key to achieving significant performance improvement or cost benefits. 
For instance, RocksDB can modularize their compaction implementation by decoupling the code logic for every primitive. 
This will not only expose the primitives as tunable knobs, but will facilitate synthesizing and testing new compaction algorithms tailored to a developer's requirements.

%Further, in the modern world of databases where change in workload specification has become a norm and performance requirements change rapidly, the observations and key takeaways presented in the paper empowers researchers/engineers with the adequate intuitions about LSM compactions. 
%Simple modifications to an existing engine, such as tuning the data movement policy or the compaction granularity, can effectively take us a long way toward extracting better performance by adapting to the changes. 

% Modern databases are often subject to workload changes and varying performance goals. 
% A common concern is that adapting to these changes can only be achieved with changes in the underlying data layout which requires extensive modification on top of the current structure. 
% However, simple modifications in the existing engine such as tuning the data movement policy or the compaction granularity can effectively take us a long way toward adapting to these changes.
% The observations and key takeaways presented in the paper empowers researchers/engineers with the adequate intuitions required for such decision-making. 

\Paragraph{Avoiding the Worst Choices}
We discuss how to avoid common pitfalls. 
For example, tiering is often considered as the 
write-optimized variant, however, we show that it comes with 
high tail latency, making it unsuitable for applications
that need worst-case performance guarantees.
Also, applications requiring stable performance should avoid \texttt{LO+2} due to its unpredictable performance.
On the other hand, partial compactions with leveling, and
especially, hybrid leveling (e.g., \texttt{1-Lvl}) offer the most
stable performance.

\Paragraph{Adapting with Workloads} 
In prior work tiering is used for write-intensive use-cases,
while leveling offers better read performance. However,
in practice, in mixed HTAP-style workloads, lookups have a 
strong temporal locality, and are essentially performed on 
recent hot data. In such cases, the block cache is frequently
proved to be enough for holding the working set and eliminate
the need for other costly optimizations for read queries.

\Paragraph{Exploring New Compaction Strategies}
Ultimately, this work lays the groundwork for exploring
the vast design space of LSM compactions.
A key intuition we developed during this analysis is that 
contrary to existing designs, LSM-based systems can benefit 
by employing different compaction primitives at different 
levels, depending on the exact workload and the performance
goals. The compaction policies we experimented with already 
support a wide range of metrics they optimize for including
system throughput, worst-case latency, read, space, and write
amplification, and delete efficiency. Using the proposed
design space, new compaction strategies can be designed with
new or combined optimization goals. We also
envision systems that automatically select 
compaction strategies on the fly depending on the current
context and workload.

% \section{Related Work}
% \label{sec:related_work}
% \input{7-related_work}

%\vspace{-0.1in}
\section{Conclusions}
\label{sec:conclusion}
LSM-based engines offer
efficient ingestion and competitive read performance,
while being able to manage various optimization goals
like write and space amplification. A key internal operation
that is at the heart of how LSM-trees work is the process
of \textit{compaction} that periodically re-organizes the data on disk.

We present the \textit{LSM compaction design space} that
uses four primitives to define compactions: (i) compaction trigger, (ii) the data layout, (iii) compaction granularity, and (iv) the data movement policy. 
We map existing approaches in this design space and
we select several representative policies to study and analyze
their impact on performance and other metrics including
write/space amplification and delete latency. We present an
extensive collection of observations, and we lay the groundwork
for LSM systems that can more flexibly navigate the design
space for compactions.

\section*{Acknowledgment}
We thank the reviewers for their valuable feedback. 
We are particularly thankful to Guanting Chen for his contributions in the early stages of this project. 
This work was partially funded by National Science Foundation under Grant No. IIS-1850202 and a Facebook Faculty Research Award.

% \appendix
% \section{Appendix}
% \label{sec:appendix}
% \input{9-appendix}

\balance 
{
% \ssmall
\bibliographystyle{abbrv}
\bibliography{/Users/subhadeep/Dropbox/W-Lab/Bibliography-Mendeley/library.bib}
 
} 

\ifx\mode\undefined
\appendix
\section*{Appendix}
\label{sec:appendix-2}

\section{Supplementary Experiments}

\begin{figure*}[!ht]
    \vspace{-1em}
         \centering
         \begin{subfigure}[b]{0.245\textwidth}
             \centering
             \includegraphics[width=\textwidth]{{exp/compaction/W1_quantile_compaction_latency_comp}.eps} 
             \vspace{-1cm}
             \caption{Overall compaction latency}
            %  \label{fig:W18_1_mean_compaction_latency}
         \end{subfigure}
         \hfill
         \begin{subfigure}[b]{0.245\textwidth}
             \centering
             \includegraphics[width=\textwidth]{{exp/compaction/W1_quantile_compaction_cpu_latency_comp}.eps}
             \vspace{-1cm}
             \caption{CPU latency for compactions}
            %  \label{fig:W18_1_write_delay}
         \end{subfigure}
         \hfill
         \begin{subfigure}[b]{0.245\textwidth}
             \centering
             \includegraphics[width=\textwidth]{{exp/point_query/W1_filter_block_cache_miss_comp}.eps}
             \vspace{-1cm}
             \caption{Block misses for point queries}
            %  \label{fig:W18_1_write_delay}
         \end{subfigure}
         \hfill
         \begin{subfigure}[b]{0.245\textwidth}
             \centering
             \includegraphics[width=\textwidth]{{exp/point_query/W1_index_block_cache_miss_comp}.eps}
             \vspace{-1cm}
             \caption{Index misses for point queries}
            %  \label{fig:W18_1_write_delay}
         \end{subfigure}
         \vspace{-0.2cm}
            \caption{(a,b) shows that correlation between the overall latency for compactions and the CPU cycles spent for compactions; (b,c) shows how the misses to the filter and index blocks change across different compaction strategies as the proportion of non-empty and empty queries change in a lookup-only workload.}
            \label{fig:W1-supp}
    \end{figure*}

In this appendix, we present the supplementary results along with the auxiliary observations (\textbf{o}) that were omitted from the main paper due to space constraints. 
In the interest of space, we limit our discussion to the most interesting results and observations. 
For better readability, we re-use the subsection titles used in \S \ref{sec:results} throughout this appendix.  

\subsection{Performance Implications}
Here we present the supplementary results for the serial execution of the ingestion-only and lookup-only workloads. 
Details about the workload specifications along with the experimental setup can be found throughout \S \ref{subsec:performance}.

\Paragraph{\mob The CPU Cost for Compactions is Significant} 
The CPU cycles spent due to compactions (Fig. \ref{fig:W1-supp}(b)) is close to $50\%$ of the overall time spent for compactions (Fig. \ref{fig:W1-supp}(a), which is same as Fig. \ref{fig:W1}(c)) regardless of the compaction strategy. 
During a compaction job CPU cycles are spent in (1) the preparation phase to obtain necessary locks and take snapshots, (2) sort-merging the entries during the compaction, (3) updating the file pointers and metadata, and (4) synchronizing the output files post compaction. 
Among these, the time spent to sort-merge the data in memory dominates the other operations. 
This explains the similarity in patterns between Fig. \ref{fig:W1-supp}(a) and \ref{fig:W1-supp}(b). 
As both the CPU time and the overall time spent for compactions are driven by the total amount of data compacted, the plots look largely similar.

\Paragraph{\mob Dissecting the Lookup Performance} 
To analyze the lookup performance presented in Fig.~\ref{fig:W1}(h), we further plot the block cache misses for Bloom filters blocks in Fig.~\ref{fig:W1-supp}(c), and the index (fence pointer) block misses in Fig.~\ref{fig:W1-supp}(d). 
Note that, both empty and non-empty lookups must first fetch the filter blocks, hence, for the filter block misses remain almost unaffected as we vary $\alpha$. 
Not that \texttt{Tier} has more misses because it has more overall sorted runs. 
Subsequently, the index blocks are fetched only if the filter probe returns positive. 
With $10$ bits-per-key the false positive is only $0.8\%$, and as we have more empty queries, that is, increasing $\alpha$, fewer index blocks are accessed. 
The filter blocks are maintained at a granularity of files and in our setup amount to $20$ I/Os. 
The index blocks are maintained for each disk page and in our setup amount to $4$ I/Os, being $1/5^{th}$ of the cost for fetching the filter blocks.\footnote{filter block size per file = \#entries per file $*$ bits-per-key = $512$*$128$*$10$B = $80$kB; index block size per file = \#entries per file $*$ (key size$+$pointer size) = $512 * (16$+$16)$B = $16$kB.}. 
The cost for fetching the filter block is $5\times$ the cost for fetching the index block. 
This, coupled with the probabilistic fetching of the index block (depending on $\alpha$ and the false positive rate ($FPR=0.8\%$) of the filter) leads to a non-monotonic latency curve for point lookups as $\alpha$ increases, and this behavior is persistent regardless of the compaction strategy.

\begin{figure}
\vspace{-1em}
     \centering
     \begin{subfigure}[b]{0.23\textwidth}
         \centering
         \includegraphics[width=\textwidth]{{exp/W18/W18.1_mean_compaction_latency_comp}.eps}
         \caption{Mean compaction latency}
         \label{fig:W18_1_mean_compaction_latency}
     \end{subfigure}
     \hfill
     \begin{subfigure}[b]{0.23\textwidth}
         \centering
         \includegraphics[width=\textwidth]{{exp/W18/W18.1_write_delay_comp}.eps}
         \caption{Write delay}
         \label{fig:W18_1_write_delay}
     \end{subfigure}
     \vspace{-1em}
        \caption{Varying Block Cache (insert-only)}
        \label{fig:W18.1}
\end{figure}

\begin{figure}
\vspace{-1em}
     \centering
     \begin{subfigure}[b]{0.23\textwidth}
         \centering
         \includegraphics[width=\textwidth]{{exp/W18/W18.2_mean_compaction_latency_comp}.eps}
         \caption{Mean compaction latency}
         \label{fig:W18_2_mean_compaction_latency}
     \end{subfigure}
     \hfill
     \begin{subfigure}[b]{0.23\textwidth}
         \centering
         \includegraphics[width=\textwidth]{{exp/W18/W18.2_write_delay_comp}.eps}
         \caption{Write delay}
         \label{fig:W18_2_write_delay}
     \end{subfigure}
     \vspace{-1em}
        \caption{Varying Block Cache (interleaving with $10\%$ point lookups)}
        \label{fig:W18.2}
\end{figure}

We vary the block cache for insert-only and mixed workloads ($10\%$ existing point lookups interleaved with insertions). For mixed workload, the mean compaction latency remains stable when block cache varies from $8$MB to $256$MB. However, for insert-only workload, the mean compaction latency increases sharply when block cache is more than $32$ MB (Fig. \ref{fig:W18_1_mean_compaction_latency} and \ref{fig:W18_2_mean_compaction_latency}). We observe that for insert-only workload, the write delay (also termed as write stall) is more than twice that of mixed workload (Fig. \ref{fig:W18_1_write_delay} and \ref{fig:W18_2_write_delay}). We leave this interesting phenomenon for future discussion. Compared to full and partial compaction, tiering is more stable with respect to different block cache size. 

\subsection{Varying Page Size}
When we vary the page size, we observe almost consist patterns across different compaction strategies for all metrics (Fig. \ref{fig:W19_1_mean_compaction_latency} and \ref{fig:W19_1_write_ampli}). It turns out that compaction strategy does not play a big role for different page sizes.

\begin{figure}
\vspace{-1em}
     \centering
     \begin{subfigure}[b]{0.23\textwidth}
         \centering
         \includegraphics[width=\textwidth]{{exp/W19/W19.1_mean_compaction_latency_comp}.eps}
         \caption{Mean compaction latency}
         \label{fig:W19_1_mean_compaction_latency}
     \end{subfigure}
     \hfill
     \begin{subfigure}[b]{0.23\textwidth}
         \centering
         \includegraphics[width=\textwidth]{{exp/W19/W19.1_write_ampli_comp}.eps}
         \caption{Write amplification}
         \label{fig:W19_1_write_ampli}
     \end{subfigure}
     \vspace{-1em}
        \caption{Varying Page Size (insert-only)}
        \label{fig:W19.1}
\end{figure}

\subsection{Varying Size Ratio}
We also compare the performance for different size ratio. According to Fig. \ref{fig:W20_1_mean_compaction_latency}, tiering has higher mean compaction latency compared to other strategies when the size ratio is no more than 6 and after 6, full compaction and oldest compaction become the top-2 time-consuming strategies. In terms of tail compaction strategy in Fig. \ref{fig:W20_1_P100_compaction_latency}, tiering is still the worst one compared to other strategies.

\begin{figure}
\vspace{-1em}
     \centering
     \begin{subfigure}[b]{0.23\textwidth}
         \centering
         \includegraphics[width=\textwidth]{{exp/compaction/W20.1_mean_compaction_latency_comp}.eps}
         \caption{Mean compaction latency}
         \label{fig:W20_1_mean_compaction_latency}
     \end{subfigure}
     \hfill
     \begin{subfigure}[b]{0.23\textwidth}
         \centering
         \includegraphics[width=\textwidth]{{exp/W20/W20.1_P100_compaction_latency_comp}.eps}
         \caption{Tail compaction latency}
         \label{fig:W20_1_P100_compaction_latency}
     \end{subfigure}
     \vspace{-1em}
        \caption{Varying Size Ratio (insert-only)}
        \label{fig:W20.1}
\end{figure}

\subsection{Varying Bits Per Key (BPK)}
We also conduct the experiment to investigate the BPK's influence over compaction.  From Fig. \ref{fig:W20_1_mean_compaction_latency} and \ref{fig:W21_2_P100_compaction_latency}, the mean and tail compaction latency may increase a little bit with increasing bits per key since larger filter blocks should be written but this increasing is very tiny since the increasing filter block is quite smaller than all data blocks. At the same, we also observe that the query latency even increases with increasing BPK (see Fig. \ref{fig:W21_2_empty_get_latency} and \ref{fig:W21_2_existing_get_latency}). This might come from higher filter block misses (Fig. YY) and this pattern becomes more obvious for existing queries in which case, accessing filter blocks is completely an extra burden. 

\begin{figure}
\vspace{-1em}
     \centering
     \begin{subfigure}[b]{0.23\textwidth}
         \centering
         \includegraphics[width=\textwidth]{{exp/W21/W21.2_mean_compaction_latency_comp}.eps}
         \caption{Mean compaction latency}
         \label{fig:W21_2_mean_compaction_latency}
     \end{subfigure}
     \hfill
     \begin{subfigure}[b]{0.23\textwidth}
         \centering
         \includegraphics[width=\textwidth]{{exp/W21/W21.2_P100_compaction_latency_comp}.eps}
         \caption{Tail compaction latency}
         \label{fig:W21_2_P100_compaction_latency}
     \end{subfigure}
     \hfill
     \begin{subfigure}[b]{0.23\textwidth}
         \centering
         \includegraphics[width=\textwidth]{{exp/W21/W21.3_empty_mean_get_latency_comp}.eps}
         \caption{Mean get latency (empty queries)}
         \label{fig:W21_2_empty_get_latency}
     \end{subfigure}
     \hfill
     \begin{subfigure}[b]{0.23\textwidth}
         \centering
         \includegraphics[width=\textwidth]{{exp/W21/W21.3_existing_mean_get_latency_comp}.eps}
         \caption{Mean get latency (existing queries)}
         \label{fig:W21_2_existing_get_latency}
     \end{subfigure}
     \vspace{-1em}
        \caption{Varying Size Ratio (insert-only)}
        \label{fig:W20.1}
\end{figure}
\fi

\end{document}